\documentclass[twocolumn]{aastex63}

\usepackage{amsmath}
\received{April 5, 2021}
\revised{April 22, 2021}
\accepted{April 29, 2021}
\published{June 22, 2021}
\shorttitle{Natural NSB during solar minimum}
\shortauthors{Alarcon, Serra-Ricart, Lemes-Perera \&  Mallorqu\'in}

\begin{document}

\title{Natural Night Sky Brightness during Solar Minimum}

\correspondingauthor{Miguel R. Alarcon}
\email{mra@iac.es}

\author[0000-0002-8134-2592]{Miguel R. Alarcon}
\affiliation{Instituto de Astrof\'\i sica de Canarias, C/Vía Láctea s/n, E-38205 La Laguna, Canarias, Spain}
\affiliation{Departamento de Astrof\'\i sica, Universidad de La Laguna, E-38205 La Laguna, Canarias, Spain}

\author[0000-0002-2394-0711]{Miquel Serra-Ricart}
\affiliation{Instituto de Astrof\'\i sica de Canarias, C/Vía Láctea s/n, E-38205 La Laguna, Canarias, Spain}
\affiliation{Departamento de Astrof\'\i sica, Universidad de La Laguna, E-38205 La Laguna, Canarias, Spain}

\author[0000-0003-1044-154X]{Samuel Lemes-Perera}
\affiliation{Instituto de Astrof\'\i sica de Canarias, C/Vía Láctea s/n, E-38205 La Laguna, Canarias, Spain}

\author[0000-0001-6394-3821]{Manuel Mallorqu\'in}
\affiliation{Instituto de Astrof\'\i sica de Canarias, C/Vía Láctea s/n, E-38205 La Laguna, Canarias, Spain}
\affiliation{Departamento de Astrof\'\i sica, Universidad de La Laguna, E-38205 La Laguna, Canarias, Spain}
\affiliation{Sieltec Canarias S.L., C/ Habitat, No. 2, Portal D, Of. 3, E-38204 La Laguna, Canarias, Spain}



\begin{abstract}
In 2018, Solar Cycle 24 entered into a solar minimum phase. During this period, 11 million zenithal night sky brightness (NSB) data were collected at different dark sites around the planet, including astronomical observatories and natural protected areas, with identical broadband Telescope Encoder and Sky Sensor photometers (based on the Unihedron Sky Quality Meter TSL237 sensor). A detailed observational review of the multiple effects that contribute to the NSB measurement has been conducted with optimal filters designed to avoid brightening effects by the Sun, the Moon, clouds, and other astronomical sources (the Galaxy and zodiacal light). The natural NSB has been calculated from the percentiles for 44 different photometers by applying these new filters. The pristine night sky was measured to change with an amplitude of 0.1 mag/arcsec$^2$ in all the photometers, which is suggested to be due to NSB variations on scales of up to months and to be compatible with semiannual oscillations. We report the systematic observation of short-time variations in NSB on the vast majority of the nights and find these to be related to airglow events forming above the mesosphere.
\end{abstract}

\keywords{Atmospheric effects (113); Night sky brightness (1112); Astronomical
instrumentation (799); Photometer (2030); Astronomical site protection (94); Observational astronomy (1145); Astronomy data analysis (1858)}


\section{Introduction} \label{sec:intro}
The natural night sky is never completely dark. Even in the
most isolated places, there is a background skyglow resulting
from both a natural component of terrestrial and extraterrestrial
origin and an artificial skyglow component resulting from
human activity. Artificial skyglow is the brightening of the
night sky due to the scattering of artificial light at night
(ALAN) by the constituents of the atmosphere (mainly gas
molecules, aerosols, and clouds) in the direction of observation.
It includes radiation that is emitted directly from the ground
and reflected upward from the surface. The loss of darkness due
to the increasing use of ALAN has a dangerous, but sometimes
neglected, impact on natural ecosystems \citep{Holker2010, Gaston2013, Bennie2016, Owens2018}. Estimates suggest that more than one-tenth of the planet’s land
area experiences ALAN and that figure rises to 23\% if
atmospheric skyglow is included \citep{Falchi2016}. Indeed,
artificially lit outdoor areas brightened at a rate of 2.2\% per
year between 2012 and 2016 \citep{Kyba2017}.\\

ALAN emissions from cities have been monitored using
instruments on Earth-orbiting satellites \citep{SanchezdeMiguel2020}. However, evaluating the effects of these emissions on night sky
brightness (NSB) in dark places (normally natural protected
areas or astronomical observatories) requires extensive groundbased
observations. NSB as seen from the ground is generated
by several sources (see \cite{RoachGordon1973} and \cite{Leinert1998} for an exhaustive overview). When the Moon is
above the horizon, its scattered light is the brightest component
of the entire night sky. On dark nights and in less polluted
places, the combination of starlight and diffuse Galactic light
(DGL) is also significant, followed by the scattering effect of
sunlight on interplanetary dust that leads to faint zodiacal light
lying around the ecliptic plane. While the stationary contribution
of these sources is linked to their reference planes,
the chemiluminescent emission of molecules above the
mesosphere, known as airglow, is not. This is not only one
of the most relevant components of the NSB in dark skies but
also a determining factor in its variability, which depends on
several parameters (season, geographical position, solar cycle,
and so on) interacting in a largely unpredictable way. Several
studies have been made to characterize night sky brightness
and its variability at the main telescope sites \citep{Walker1988, Leinert1995, Mattila1996, Krisciunas1997, Benn1998, Patat2003, Patat2008}. However, a regular and systematic study is necessary to monitor possible trends toward
the brightening of the global natural NSB and to investigate the
different processes driving its variability.\\

The analysis presented here is based on zenith NSB
measurements performed by a night photometer network
deployed by the European project STARS4ALL \citep{Zamorano2017}. Zenithal NSB, despite having considerably limited information content than all-sky or photometric systems, is widely used to characterize NSB. This is partly
due to the widespread availability of low-cost detectors that
enable its straightforward measurement by professional and
citizen scientists worldwide \citep{Alejandro2017, Posch2018, Puschnig2020}. This work aims to
define the first comprehensive reference method to combine
measurements from several low-cost broadband photometers
distributed in different locations for the study of natural night
sky darkness. This will allow us first to determine the intensity
of the local skyglow and then to study its variability and
possible origin by detecting the changes that the artificial
disruption of night sky darkness may experience in the coming
years. This is a relevant issue not only for preservation of
the capabilities of observational astronomical sites \cite{Walker1970} but also for biodiversity preservation, urban emission monitoring \citep{Bustamante2021}, energy economics \citep{Gallaway2010}, and preservation of the night sky as a key asset of the intangible cultural heritage of humanity.\\

The paper is organized as follows. The different detectors
used in data acquisition are described in Section \ref{sec:data}; the method
of filtering data with different constraints (e.g., clouds, the
Moon, and Galactic and zodiacal light) is discussed in Section \ref{sec:analysis}, and the results obtained during dark time are then
presented in Section \ref{sec:naturalsky}. Timescale variations and their origin are
analyzed in Section \ref{sec:discu} and the main conclusions are presented in
Section \ref{sec:conclu}. Finally, detailed discussions about the photometric,
calibration, and instrumental uncertainties of the Telescope
Encoder and Sky Sensor (TESS) photometer are given in the Appendix.

\section{Data Acquisition} \label{sec:data}
\subsection{TESS Photometer and Sky Quality Meter} \label{subsec:tess}
The European-funded project STARS4ALL\footnote{\url{www.stars4all.eu}} has recently developed a new detector that uses the same TSL237 photodiode detector as the Sky Quality Meter (SQM; \citet{Cinzano2007}). It is a silicon photodiode with a light-to-frequency
converter, which has a spectral response that varies
from 300 to 1100 nm approximately, with a sensitivity
maximum around 700 nm, responding linearly to frequency
over several irradiance orders \citep{Bar2019}. This
instrument is called TESS \citet{Zamorano2017}). The
SQM uses a Hoya CM-500 filter for nominally limiting its
effective bandpass to 400-–650 nm, whereas TESS is fitted with
a dichroic filter that limits it to the 400-–740 nm spectral band,
allowing for better coverage of the red band of the visible
spectrum with good response (the spectral response of these
photometers is shown in the Appendix). A concentrator optic
restricts the device field of view (FOV) to a region of the sky
with an approximately Gaussian transmittance profile with an
FWHM of 17$^\circ$.\\

TESS was designed with the goal of creating a large
European network with inexpensive but well-tested photometers.
In fact, the device is calibrated by the manufacturer at
the Laboratory for Scientific Advanced Instrumentation
(Laboratorio de Instrumentación Científica Avanzada) of
Universidad Complutense de Madrid (UCM, Spain) and
contains a complete system to transmit data to STARS4ALL
servers. At present more than 200 photometers are installed and
sending data worldwide, which are available at the project
website\footnote{\url{tess.dashboards.stars4all.eu}}. To avoid mistakes, it is necessary to consider the
TESS and SQM responses as those of different photometric
systems. The conversion factors between TESS, the SQM, and
other photometric systems can be obtained based on the kind of
spectrum of the observed object \citep{Bar2019}. Hereafter
the brightness in magnitude per square arcsecond measured in
the TESS and SQM passbands will be called $m_{\rm TESS}$ and $m_{\rm SQM}$,
respectively. A total of 44 TESS photometers (see Table \ref{table-TESS}) and
11 million individual raw measurements have been included in
the present work. A detailed analysis of the uncertainties
existing in the photometer data and the differential photometry
is given in the Appendix.

\subsection{ASTMON all-sky camera} \label{subsec:ASTMON}
The All-sky Transmission Monitor (ASTMON) instrument
is designed to perform continuous monitoring of NSB in
several bands \citep{Aceituno2011}. At present two ASTMON
stations are installed at Observatorio del Roque de los
Muchachos (ORM, La Palma) and a third one at Observatorio
del Teide (OT, Tenerife) in the Canary Islands, all operated by
Instituto de Astrofísica de Canarias (IAC, Spain). All are fully
robotic and have standard Johnson–Cousins \textit{BVRI} filters \citep{Bessell1990}. 

\subsection{Light pollution laboratory} \label{subsec:IOT-EELab}
The measurements gathered by the STARS4ALL network
are openly available in real time for researchers and the public
at the Internet of Things EELabs (IOT-EELab) website\footnote{\url{data.eelabs.eu}}. IOT-EELab
is an interactive website dashboard that collects,
controls, and analyzes data from hundreds of internetconnected
light pollution devices, including night photometers,
LED control sensors, and all-sky cameras.
IOT-EELab provides a unique opportunity for researchers to
experiment with new and existing light pollution data sets to
apply data science and big data techniques to gain new insights
and valuable information, or build their own applications or
research projects.

\section{Data analysis} \label{sec:analysis}
NSB as seen from the ground is generated by several layers
at vastly different distances. Neglecting the small contribution
from extragalactic background light (which, at visible wavelengths,
represents only 0.2\% of the total NSB; see Table 4 of \citet{Masana2021}), the main contributors to NSB are
integrated starlight (ISL) and diffuse background radiation
(both originating in our Galaxy); zodiacal light, caused by the
scattering of solar radiation from interplanetary dust; airglow
emission from high atmospheric layers; and the scattering of
these components in the troposphere \citep{Leinert1998}. In
addition to these natural components, human activity has added
an extra source, namely the artificial light emitted upward from
cities and scattered back by the troposphere. New analytical
techniques that select those data points that are uninfluenced by
the Sun, the Moon, clouds, and astronomical sources (Galactic
and zodiacal light) are presented in the following sections.

\subsection{Cloud coverage analysis} \label{subsec:cloud}
One of the main natural factors that affect sky brightness
measurements is the presence of clouds in the photometer’s
FOV. When illuminated by either natural (the Moon) or
artificial (ALAN) external sources, clouds produce a significant
increase in the skyglow of urban sites \citep{Kyba2011, Jechow2017}, whereas the opposite occurs in dark areas \citep{Jechow2019}. Therefore, defining a robust method so it
has only cloudless nights and knowing the uncertainty
associated with them is essential in order to properly
characterize the NSB. In this work, the performance of two
different methods is tested: one based on the difference
between the ambient and IR temperatures of the sky, obtained
with the thermometers incorporated in TESS, and another
based on the high variability in the measurement of NSB in the
presence of clouds given by the standard deviation \citep{Cavazzani2020}.\\

\begin{figure}[t]
\begin{center}
\includegraphics[width=\linewidth]{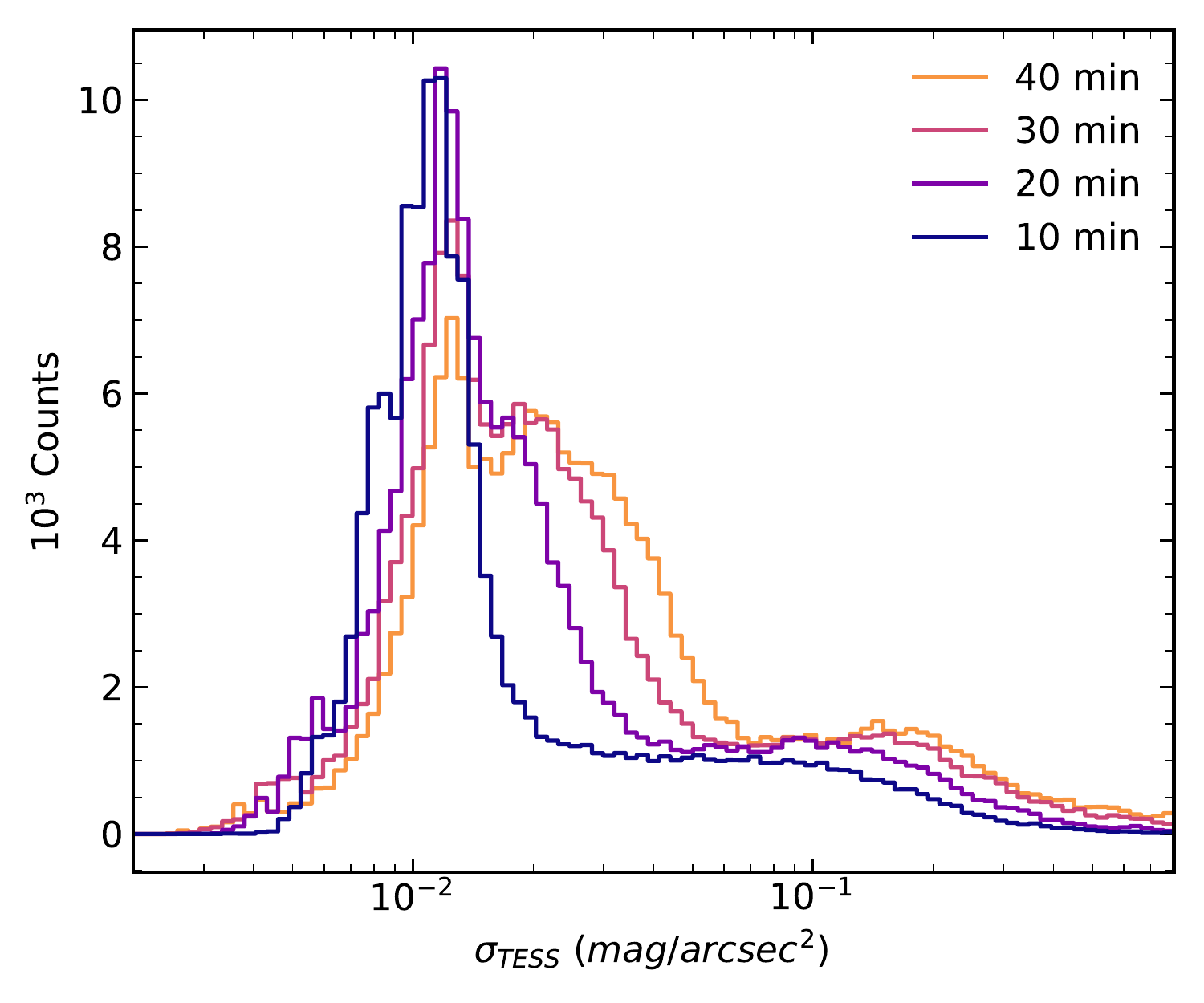}
\caption{Standard deviation $\sigma_{\rm TESS}$ of NSB calculated over different time
periods. Each distribution has ${\sim}$200,000 data, measured with the photometer stars211 located at the Observatorio del Teide (Tenerife, Canary Islands, Spain). Data affected by the Sun and Moon have been filtered out. \label{fig:cloud_hist}}
\end{center}
\end{figure}

While in the former method measurements of NSB and the
ambient and IR temperatures are taken simultaneously, in the
latter it is mandatory to define a period just before the reading over which the standard deviation is calculated. Choosing the
length of the period is not trivial, as the $\sigma_{\rm TESS}$ distribution tends to spread out as it gets longer --see Figure \ref{fig:cloud_hist}. This is a
consequence of the many factors involved in the short-term
variability of sky brightness that are analyzed in this work.
Both the value of $\sigma_{\rm TESS}$ and the length of the period must be
fine-tuned to filter out the cloudy data.\\

To test the performance of both methods, ${\sim}$4000 all-sky
images taken with All-sky Meteor Orbit System (AMOS)
cameras \citep{AMOS2013} at OT and ORM during 2019 have
been labeled visually. These high-sensitivity cameras designed
for meteor detection make the presence of thin cirrus clouds
distinguishable in the FOV covered by TESS. There is a bias of
more meteors detected on clear nights, which represent three
quarters of the whole data set. To balance out this proportion,
overcast nights are added. At OT there are two cloud (Song and
Open University Telescopes) and four humidity detectors
(Song, Open University, STELLA Telescopes, and Residence).
The sky is considered completely overcast when both cloud
detectors are positive and humidity is 100\% in all four
detectors. For each method and period, the measurements of the
photometers stars211 (OT) and stars90 (ORM) are merged with
the AMOS images and inspected. Cloudlessness is deemed to
occur when all the images within the period have been
classified as clear, conditions otherwise being considered
cloudy. In total, between 33,000 (for a period of 40 minutes)
and 15,000 (10 minutes) TESS measurements unaffected by
sunlight and moonlight have been annotated as clear/cloudy
and used to test the methods.\\

\begin{figure}[t]
\begin{center}
\includegraphics[width=\linewidth]{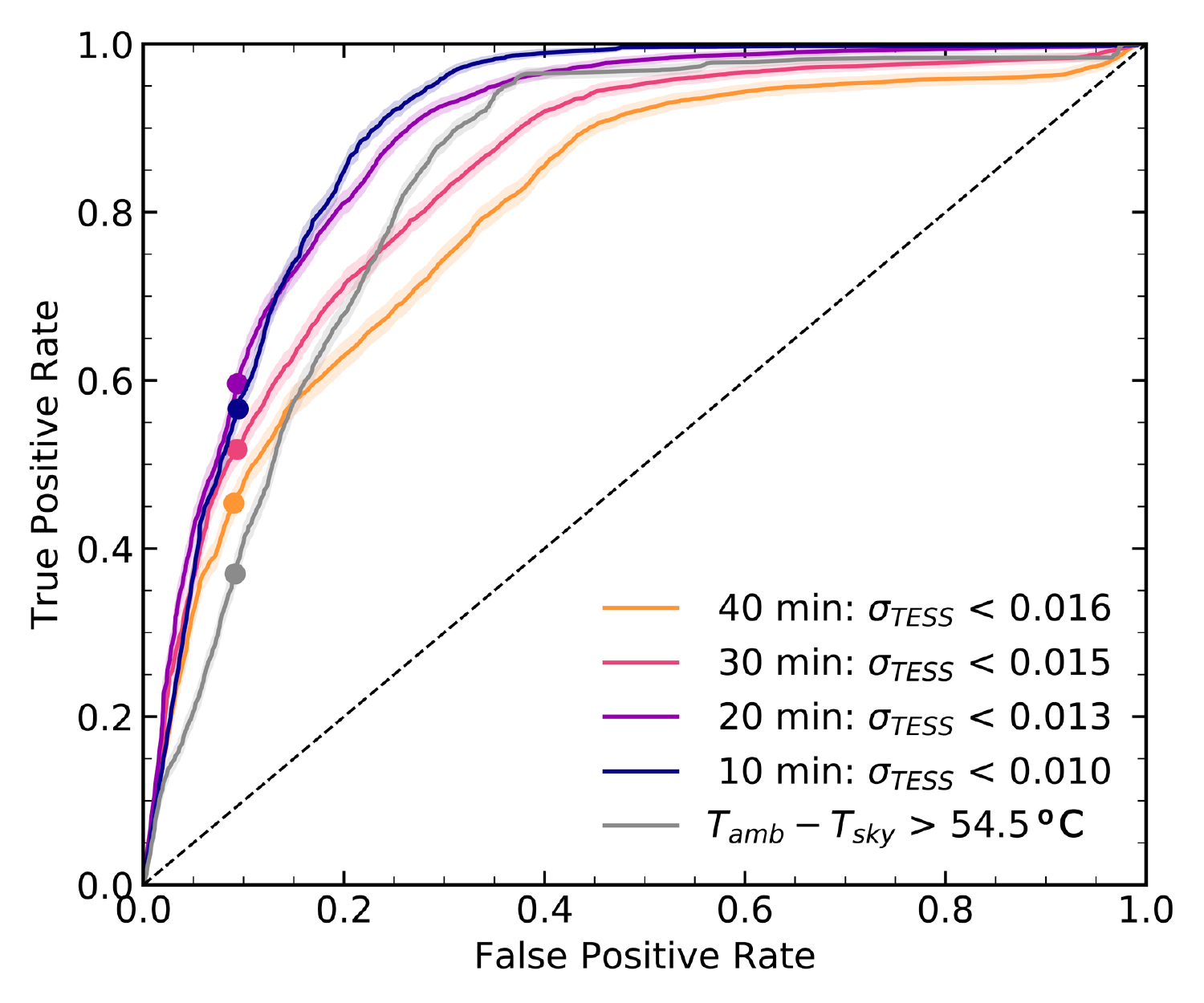}
\caption{ROC curves for the standard deviation method $\sigma_{\rm TESS}$, calculated for
different periods, and the temperature difference method (black). The dots
indicate the points where the criterion FPR = 0.1 is satisfied, whose threshold
for the data to be considered cloudless is indicated in the legend. The diagonal dashed line is the reference for a random classifier.\label{fig:roc}}
\end{center}
\end{figure}

To fine-tune the threshold to below where the sky is
considered cloudless, the receiver operating characteristic
(ROC) curve \cite{Swets1988}) has been obtained for both methods.
It represents the true-positive rate (TPR) against the false-positive
rate (FPR) for a range of threshold values, and can be
used to compare different models, as shown in Figure \ref{fig:roc}. A
classifier is considered to be good when it maximizes the TPR,
losing less clear data, while minimizing the FPR, reducing the
probability of including partially/totally cloudy data as clear.
To avoid possible deviations due to different data set sizes and
to estimate the uncertainty in the TPR and FPR, the data set has
been bootstrapped by taking 1000 random samples of 5000
data with replacements and calculating their median and 95\%
confidence interval, which are shown in the figure as a thick
line and band, respectively.\\

There is no universal criterion for choosing the threshold.
Although the closest to the top left corner is usually taken as
the best value, this may depend on the specific conditions of the
problem. We are interested in filtering out data with some cloud
presence as far as possible, even if this means losing a lot of
clear data. The authors consider that an FPR of 10\% is
reasonable for the study of natural NSB with these low-cost
photometers, but other equally valid values could be chosen for
other types of studies, interests, or instruments. The points that
meet this criterion are shown on the ROC curves and the
threshold is included in the legend. It should be noted that the
$\sigma_{\rm TESS}$ method with a period of 10 and 20 minutes is better than
the temperature difference method--which is used by default in
the TESS photometers--over the whole range. Both intervals
are similar in the low-FPR region and even though for 20
minutes the FPR is a little bit higher than that for 10 minutes,
the latter is considered to be the better one for our purposes.
First, the shorter the period, the smaller, on average, the
difference between the measurement and the labeled data
timestamp being compared, thus increasing the reliability of the
true-condition data set. On the other hand, the global
performance of a classifier can be estimated by the area under
the curve (AUC), considered better if it is closer to 1--closer to
the upper left corner--or to 0.5 for a random choice (the
diagonal dashed line in the figure). The highest AUC is
obtained by taking a 10 minute interval, 0.892$\pm$0.005,
followed by a 20 minute interval, 0.881$\pm$0.006, and the
temperature difference, 0.837$\pm$0.007. Hence, cloudlessness is
considered to occur when the standard deviation of the sky
brightness measurements in the preceding 10 minutes is
$\sigma_{\rm TESS} < 0.010$ mag/arcsec$^2$, with a TPR of 0.566$\pm$0.015
and an FPR of 0.094$\pm$0.006.

At this point, it should be noted that these curves and
thresholds may be different for brighter locations, as there is a
dependency of the standard deviation on NSB, as noted by \cite{Cavazzani2020}. Our analysis is considered valid for the
range of magnitudes studied in this work. Other methods such
as the IR temperature or sky brightness derivative method
proposed by \cite{Tomasz2020} have been tested without better
results.\\

One of the greatest advantages of knowing the TPR and FPR
is being able to estimate correctly the percentage of clear night
time on every photometer. If this were calculated by means of
the fraction of data that fulfills the condition of $\sigma_{\rm TESS}$ against
the total, the result would be highly biased, because it would
include $\sim$10\% of false positives and discard $\sim$43\% of true
positives. To correct this bias we use the Rogan--Gladen
estimator \citep{Rogan1978}:
\begin{equation}
    P_{\rm corr} = \frac{P - FPR}{TPR - FPR}
\end{equation}
where $P$ and $P_{\rm corr}$ are the percentages of clear time before and
after correction. The result for each photometer used in this
paper is included in Table \ref{table-TESS}. The uncertainties in these values
are estimated from the bootstrapped TPR and FPR ranges.

\subsection{Astronomical components} \label{subsec:astro-com}
In the following sections, we will use the combined
information from all the available photometers to characterize
the effects of the different astronomical components affecting
the natural NSB. To do this, it is essential to correct for
differences in brightness measurements at the different sites. To
find the offset for each photometer we have taken the median
value of its natural NSB, P50$_{\rm TESS}$, as a reference, calculated
iteratively from the clear data remaining after applying all the
filters that will be discussed below. By subtracting the P50
values of each photometer from its magnitude measurements,
they can be treated as independent samples as long as they are
uncorrelated—this will be discussed extensively in Section \ref{subsec:short-term}. Hence, the variations in NSB that will be discussed hereafter have been characterized from the mean of the $m_{\rm TESS}$--P50$_{\rm TESS}$ of all the photometers and their standard errors.
 
\subsubsection{Sun}\label{sec:sunlight}
Once the Sun goes below the horizon and civil twilight
begins, the sky brightness starts to decrease quickly. As the
solar elevation decreases, the atmospheric layers directly
illuminated by the Sun become higher and the blue part of
the zenith spectrum becomes stronger as a result of ozone
absorption until it reaches its maximum at the end of nautical
twilight (12$^\circ$ below the horizon). From that moment on, when
many of the brightest stars are already visible, the sky spectrum
and intensity profile begin to flatten out, becoming gradually
redder and darker until the end of astronomical twilight, when
the elevation of the Sun is $-18^\circ$ \citep{Spitschan2016}. The
reverse process occurs at dawn. In this study, it is considered,
as is usually done in observational astronomy, that the NSB
measurements are not affected by direct sunlight when the Sun
is below $-18^\circ$ (see, for example, Figures 3--7 in \cite{Patat2006}). Other indirect effects related to solar radiation will be
described below.

\subsubsection{Moonlight}\label{sec:moonlight}
The Moon is the major contributor to natural light pollution
at night, especially in the bluest part of the spectrum, because
of the efficiency of Rayleigh/Mie scattering increasing toward
shorter wavelengths \citep{Noll2012}. NSB due to scattered
moonlight depends on many factors. First, the intensity of light
that enters our atmosphere after being reflected by the lunar
surface is a function of its albedo and distance. The albedo
depends primarily on the phase, but also on the solar
selenographic longitude and wavelength. The variation of the
distance to the Moon can affect the incident intensity in the
atmosphere by 10\% at maximum \citep{Jones2013}. In the
atmosphere, moonlight is affected by Rayleigh and Mie
scattering, which depend mainly on the transmission of the
atmosphere for the airmass of the Moon and the target, their
angular separation, and their wavelength.\\

\begin{figure}[t]
\begin{center}
\includegraphics[width=\linewidth]{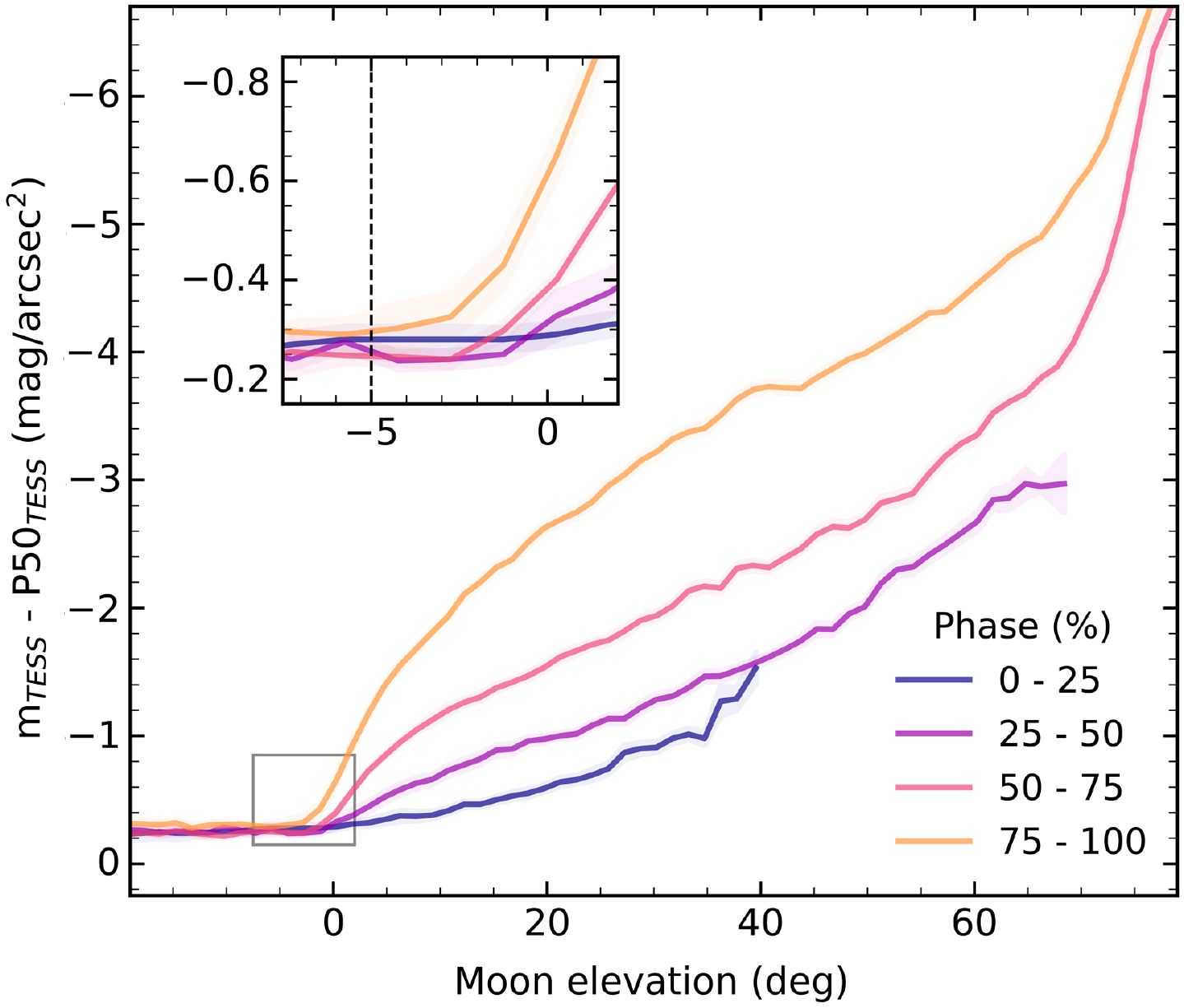}
\caption{NSB difference against Moon elevation for different lunar phases.
The region near the horizon is zoomed in on the inside plot, indicating the
elevation (vertical dashed line) from which the data are filtered. All available photometers have been included, making a total of 8.13 million measurements.\label{fig:moon}}
\end{center}
\end{figure}

The fact that all our measurements have been taken at the
zenith makes the problem much easier, especially in the
treatment of scattering. The intensity of moonlight can be
parameterized according to its phase--or illumination--and
elevation. Thus, we will analyze how NSB changes according
to these two parameters. As a reminder, the aim of this work
is not to make an empirical model of the different
components that affect NSB, but to establish filters that
allow us to minimize them to study the natural NSB. Figure \ref{fig:moon} shows the NSB difference as a function of the elevation and
phase of the Moon for all the photometers, filtering data
affected by sunlight. The brightness curves remain constant
for Moon elevations below $-5^\circ$,regardless of phase. Above
this value, it is considered that the measurements could be
polluted by moonlight and should be filtered out. It should be
noted that the $m_{\rm TESS}$--P50$_{\rm TESS}$ difference when the Moon is
below the horizon does not reach 0 because of the presence of
other bright components such as the Galaxy and zodiacal
light.

\subsubsection{Galaxy}\label{sec:galaxy}
The Milky Way, zodiacal light, and airglow are the most
important components affecting NSB on clear, moonless
astronomical (Sun elevation $<-18^\circ$) nights in the visible
spectral range. When observed very close to the Galactic plane,
the contribution of starlight (ISL) and diffuse background
radiation (DGL), which can be 20--30\% of that integrated
into the FOV of the photometer, may be the greatest of all \citep{Leinert1998}. This is especially the case at midlatitudes
in the southern hemisphere, where the Galactic center is close
to the zenith in fall and winter. Decoupling these three
components in ground-based observations to study them
separately is not trivial. The data taken by the imaging
photopolarimeters on board Pioneer 10 and 11 during their
journey beyond the outer asteroid belt are currently the main
reference for determining the absolute brightness of the Galaxy
(see, for example, \cite{Mellinger2009}, \cite{Zou2010}, \cite{Noll2012}), because only in those regions of the solar system ($R>2.8$ AU) is the contribution of zodiacal light negligible \citep{Weinberg1974,Schuerman1977}. Any other ground- or space-based ISL+DGL measurements must be
either complemented by semiempirical models to correct for
these contributions (e.g., \cite{Hoffmann1998, Duriscoe2013, Gill2020}) or constrained to specific observing conditions,
which is the approach of this work.\\

To filter out the contribution of Galactic light, a visual
inspection of the data from all the photometers, once sunlight,
moonlight, and clouds have been filtered out, has been
conducted and some common patterns have been identified.
In the Galactic plane, the central region is clearly the brightest,
with a rapid decrease toward the arms and smoother with
increasing latitude. With such a wide FOV, inhomogeneities
produced by dark clouds or star clusters in the Galactic plane
are not expected to be detected, but high symmetry in both
longitude and latitude is. It might be expected that, regardless
of the Galactic longitude, the NSB difference due to the Galaxy
would decrease at higher latitudes until it is almost residual.
However, there are two regions where this does not apply:
the center and the anticenter, where the Galactic and ecliptic
planes are overlapped. Hence, a distinction has been made
as to whether zodiacal light could be shifting the curves
toward brighter values, according to the method discussed in
Section \ref{sec:zodiacal}.\\

\begin{figure}[t]
\begin{center}
\includegraphics[width=\linewidth]{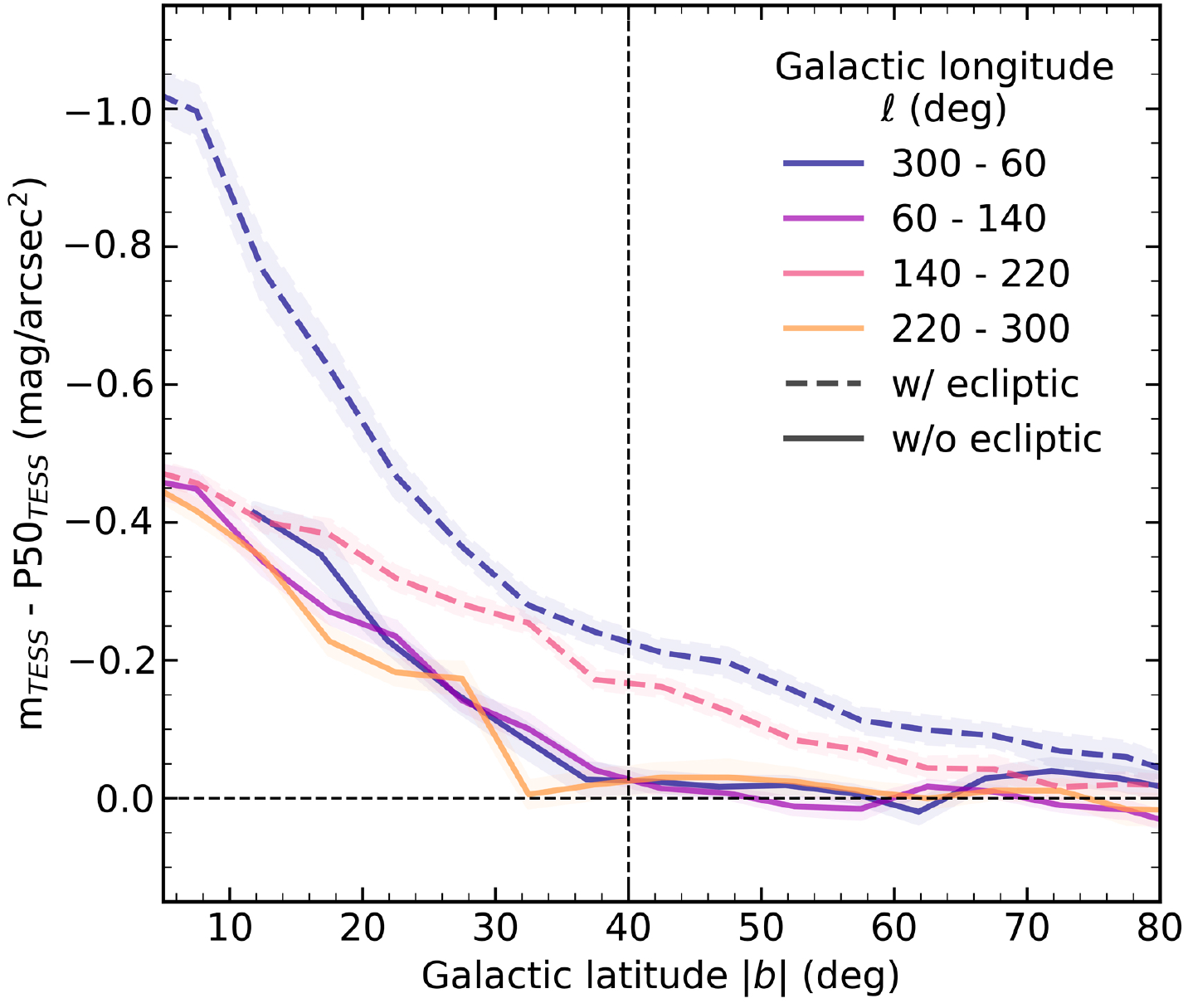}
\caption{NSB difference against absolute Galactic latitude for different
longitude intervals covering the center (blue), anticenter (pink), and arms
(yellow and purple) of the Galaxy. Measurements that can be affected by
zodiacal light according to Section \ref{sec:zodiacal} are shown as dashed lines, while the
rest are shown as solid lines. To avoid the contribution of the Galaxy in the
clean data, those with $|b| < 40^\circ$ are filtered out, indicated by a vertical black
dashed line. All available photometers have been included, making a total of
1.44 million data on moonless and cloudless astronomical nights.\label{fig:gal}}
\end{center}
\end{figure}

The NSB difference as a function of the absolute value of the
Galactic latitude for different longitudes is shown in Figure \ref{fig:gal}. In the region close to the Galactic center (blue line) the zenith
sky brightness can increase by up to 1 mag/arcsec$^2$, while the
average elsewhere in the Galactic plane reaches about 0.4 mag/arcsec$^2$. The effect of the ecliptic is also shown in the figure.
While in the regions affected by zodiacal light the NSB
difference does not reach 0 up to very high Galactic latitudes
near the poles, the solid curves, corresponding to measurements
far from the ecliptic, flatten out for $|b|\geq40^\circ$. Therefore, in this
work it will be considered that measurements are unaffected by
Galactic light when the absolute value of the Galactic latitude
at zenith is greater than $40^\circ$.

\subsubsection{Zodiacal light}\label{sec:zodiacal}
Zodiacal light is the result of the scattering of sunlight by
interplanetary dust particles in the solar system. Its intensity
depends mainly on ecliptic latitude $\beta$ and helioecliptic
longitude $\lambda-\lambda_\odot$, defined as the difference between the
longitudes of the object, in our case the zenith, and the Sun.
In contrast to the Galactic component, the brightness map of
zodiacal light is not fixed to the celestial reference system, but
depends on the position of the Sun in the sky and therefore on
the location and time of year. As the Sun goes below the
horizon and moves along the ecliptic, the brightness smoothly
decreases until it reaches longitudes close to $180^\circ$, where it
increases steeply in a small area of about $20^\circ$ in diameter
known as the Gegenschein \citep{Leinert1998}.\\

Several brightness maps have been used in the literature to
take into account the contribution of zodiacal light. For
instance, \cite{Patat2003} interpolated the data presented by \cite{LevasseurDumont1980} to correct for zodiacal
light in his observations from Cerro Paranal. An improved
model made by \cite{Leinert1998} was used by \cite{Noll2012} and \cite{Masana2021} to account for reddening and
thermal emission. Likewise, the model obtained by \cite{Kwon2004} from photopolarimetric data taken on consecutive nights
from Mount Haleakala in Hawaii was used as a reference by \cite{Duriscoe2013} and \cite{Gill2020} to quantify brightness
excess due to zodiacal light. Right from the outset of this work,
we have avoided doing any model-derived correction in order
to directly process the raw data.\\

\begin{figure}[t]
\begin{center}
\includegraphics[width=\linewidth]{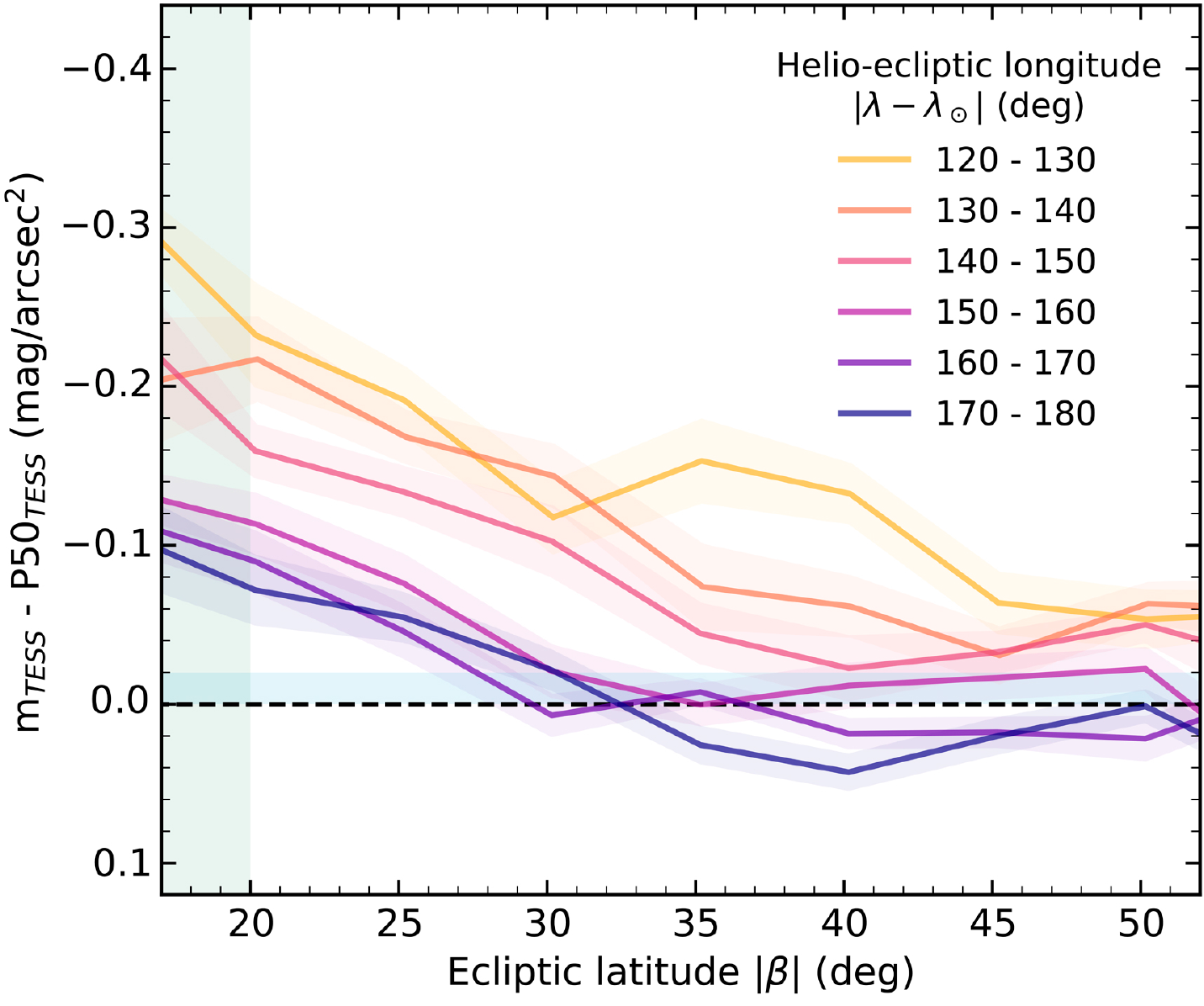}
\caption{NSB difference against absolute ecliptic latitude for different
helioecliptic longitudes, defined as the difference between the solar and
zenithal ecliptic longitude for our specific observing site. The horizontal dashed
line represents the absence of zodiacal light affecting the natural NSB
measurements. The relationship between latitude and longitude in this blue-colored
area is shown in Figure \ref{fig:ecl_extra}a. The vicinity of the ecliptic plane,
indicated by a vertical green band, is shown enlarged in Figure \ref{fig:ecl_extra}b.\label{fig:ecl}}
\end{center}
\end{figure}

\begin{figure}[t]
\gridline{\fig{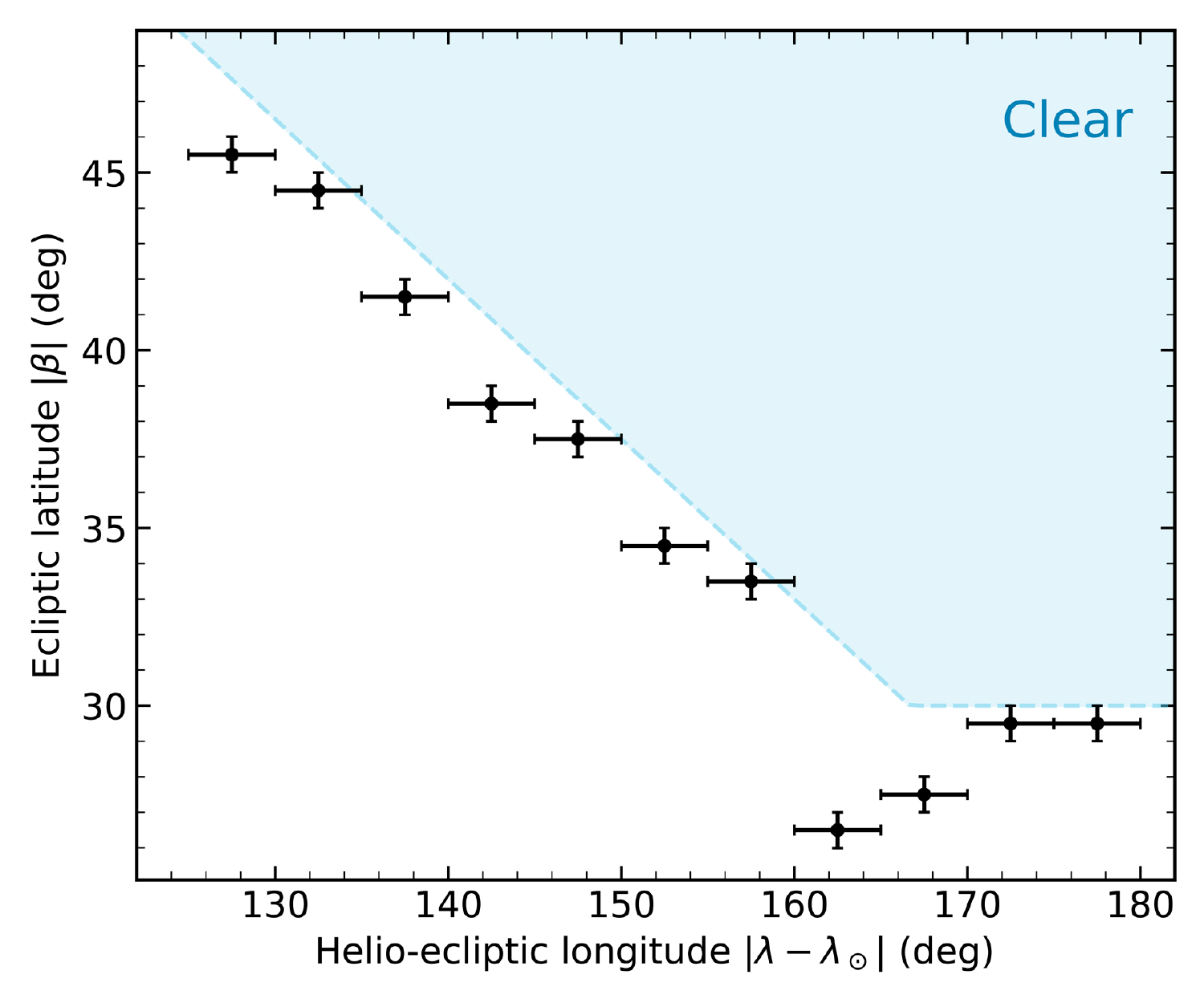}{\linewidth}{(a) Relation between the helio-ecliptic longitude and latitude of the points where the condition m$_{\rm TESS}$--P50$_{\rm TESS} = 0$ is satisfied and the contribution of the zodiacal light is considered to be negligible. The corresponding filter is defined by the blue dashed line. Thus, all points in the blue area are considered to be clear of astronomical components.}}
\gridline{\fig{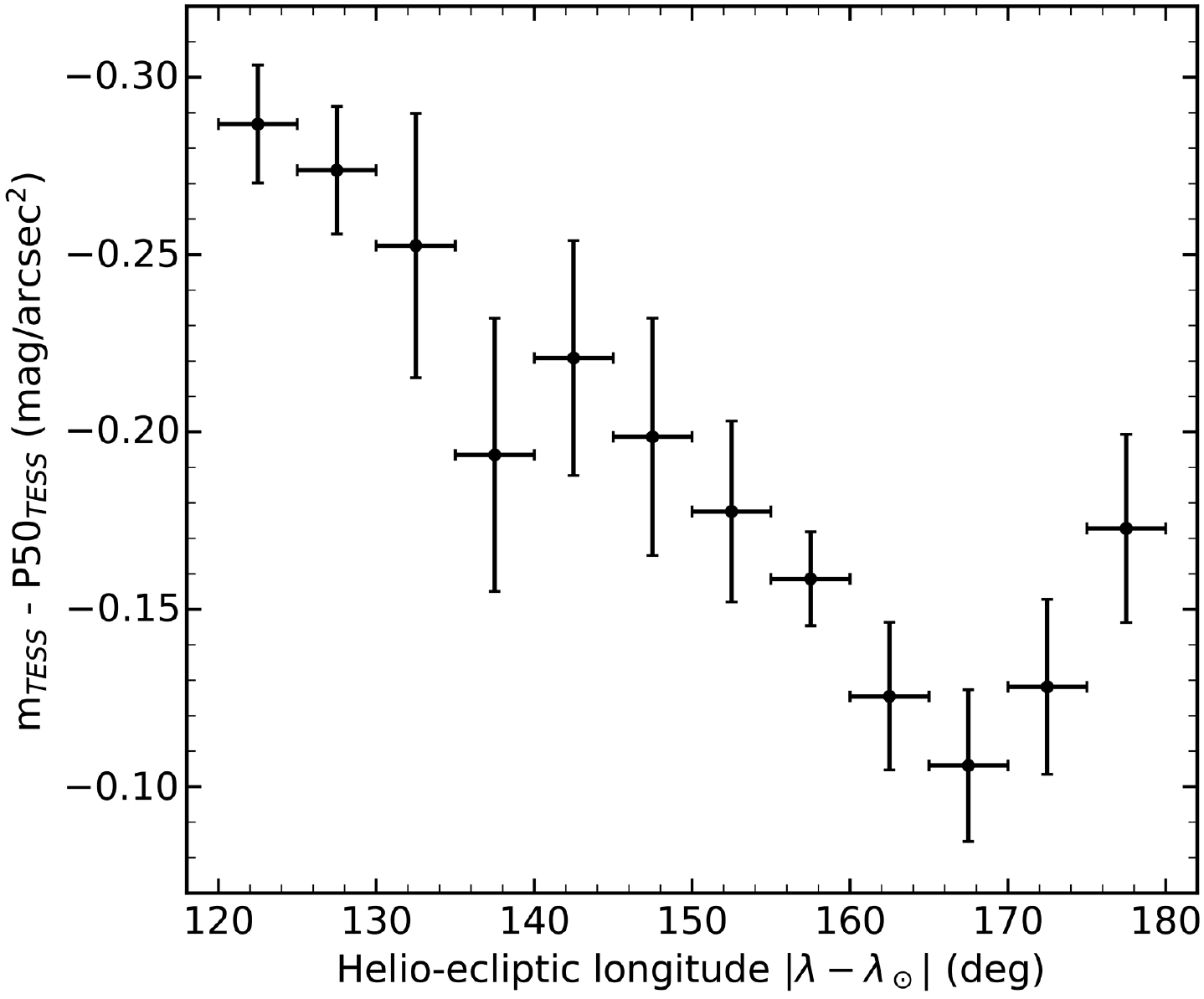}{\linewidth}{(b) NSB difference versus helio-ecliptic longitude for zenithal observations near the ecliptic, i.e.\ with latitude $|\beta|<20$. The brightness bump near the antisolar point corresponds to the Gegenschein.}}
\caption{Detail of the border region in ecliptic coordinates where the zodiacal light filter is defined (top), marked with a blue horizontal band in Figure \ref{fig:ecl}, and the vicinity of the ecliptic (bottom), marked with a green vertical band in 
Figure \ref{fig:ecl}. Each black dot corresponds to a bin of $5^\circ$ in helio-ecliptic longitudes.\label{fig:ecl_extra}}
\end{figure}

After filtering out the effect of the Sun, Moon, clouds, and
Galaxy, there are still 520,000 data left with which to study the
contribution of zodiacal light. To avoid possible post-twilight
brightness enhancements, which are analyzed in Section \ref{subsec:walker}, all
measurements corresponding to the first 4 hr of the night
have been removed. The NSB difference as a function of
helioecliptic longitude and latitude is presented in Figure \ref{fig:ecl}. As
expected, there is a clear trend toward a fading of zodiacal light
with increasing separation from the Sun and the ecliptic. We
search for the ecliptic latitude at which the difference between $m_{\rm TESS}$ and P50$_{\rm TESS}$ is near 0, indicated in the figure by a horizontal black dashed line. This is equivalent to finding the
cutoff points between the horizontal blue band in the figure and
each of the curves. All points above this band are considered to
be unaffected by zodiacal light. The procedure, illustrated in Figure \ref{fig:ecl_extra}a, is as follows.\\

We restrict our data to the 28 darker photometers, corresponding
to a P50$_{\rm TESS} > 21.5$ mag/arcsec$^2$, and define a
longitude bin size of $5^\circ$, which is wide enough to avoid
fluctuations between the data as a consequence of having a
small sample (around 20,000 measurements each), but still
allow a good sampling of the latitude--longitude curve. For
each bin, we seek the value of the ecliptic latitude for which the
sky brightness difference is greater than 0, i.e., for which the
zodiacal light effect is negligible. This value of ecliptic latitude
for each helioecliptic longitude bin is represented in Figure \ref{fig:ecl_extra}a as a black dot. As we increase separation from the Sun, the
ecliptic latitude from which zodiacal light begins to fade
becomes smaller. This is what can be seen in brightness maps
such as those made by \cite{Kwon2004}, with a broader
distribution near the horizon at dusk and dawn that gradually
narrows toward the antisolar point. The only exception is near
this position, where the Gegenschein can be noticed. Indeed, Figure \ref{fig:ecl_extra}b shows the sky brightness difference as a function of
helioecliptic longitude by adding up all the data for ecliptic
latitude less than $20^\circ$ (the vertical green band in Figure \ref{fig:ecl}). The
sudden increase in brightness from $165^\circ$ onward should be
noted; this is indicative of the detection of the Gegenschein
with the TESS photometers. It is expected that for all points
above the black ones in Figure \ref{fig:ecl_extra}a the contribution of zodiacal
light is negligible. Hence, these points are used as a reference to
set the ecliptic filter:
\begin{equation}
|\beta| > {\rm max}\left(105 - 0.45\cdot |\lambda-\lambda_\odot|, 30\right) \quad ({\rm deg}) \label{eq:ecl}
\end{equation}
where the maximum condition is introduced to take into
account the Gegenschein enhancement. This expression defines
the set of ecliptic coordinates of the clear points, which is
shown as a blue-highlighted area in Figure \ref{fig:ecl_extra}a.

\subsection{Atmospheric components}
\subsubsection{Natural airglow} \label{subsec:airglow}
High-energy solar radiation incident on the upper atmosphere,
around 90–100 km (mesopause) and 270 km (ionospheric
F2 layer), leads to numerous photochemical processes
that result in light emission known as airglow. These
chemoluminescence processes are the result of the decay of
excited states of atoms and molecules produced by chemical
reactions (see \cite{Khomich2008} for a detailed description).
In the visible band, airglow is characterized by a multicomponent
pseudo-continuum with reactions of iron and nitric
oxide, atomic oxygen, or ozone and several emission lines that
dominate the night sky spectrum. For instance, the relaxation of
atomic oxygen previously excited to the $^1$S state in a two-step
process involving O$_2$, results in the spontaneous emission of a $5577$ $\r{A}$ photon \citep{Barth1961}, which is the most
prominent line and gives the airglow its characteristic green
color. On the other hand, the common red color is mainly
due to the forbidden-line emission by the dissociative recombination
of [OI] $6300$ in a higher layer of the atmosphere
(250--300 km) and the relaxation of vibrationally excited OH
radicals resulting from the ozone hydration process \citep{Meinel1950}. Other lines, such as those of the Na I D doublet or
molecular oxygen, can also be seen in the visible spectrum \citep{RoachGordon1973}, some of them included in Figure \ref{fig:spectra}.\\

While the astronomical components described in the
previous section could be modeled quite accurately owing to
their stationarity in their respective reference systems, airglow
is not only highly variable but also difficult to predict. The
brightness of airglow lines can vary by up to 20\% in tens of
minutes over a long temporal and spatial range. The numerous
reactions that result in airglow can be affected by multiple
factors, including solar activity, seasonal trends, changes in
atmospheric variables or the composition of the upper layers,
geomagnetic activity, gravity waves, and so forth \citep{Khomich2008}. Such a complex system makes it difficult to
perform systematic processing of photometric data, so no filter
is defined to model this. Moreover, airglow is considered an
important element in natural NSB and will be discussed in the
next sections.

\subsubsection{Artificial skyglow} \label{subsec:skyglow}
A non-negligible part of artificial outdoor lighting is emitted
toward the sky directly or reflected from different surfaces,
resulting in an increase in sky brightness known as skyglow \citep{Falchi2016}. Although changes in ground-based source
emissions modulated by atmospheric scattering processes in the
presence of dust particles, water droplets, and other atmospheric
components can be reflected in the spectral properties
of the diffuse light (e.g., \cite{Benn1998}), their
counterpart in broadband photometers is not so obvious \citep{Alejandro2017}. Particularly in dark locations
with low outdoor lighting density, artificial skyglow is not
expected to show short-term variations, but a stationary
intensity, characteristic of each location, which constitutes the
main driver of its darkness.

There is a possibility that NSB measurements show an
abrupt change as a consequence of some modification in the
nearby illumination that directly affects it. This is the case, for
instance, of the photometer stars271, located in Puerto Villareal
(Extremadura, Spain), whose data show a sudden brightening
at the end of 2019 as a consequence of a new streetlight
installed a few meters away from the device. The method
described in this section permits not only the determination of
the darkness of the site but also the identification of possible
changes in the observational configuration from the distribution
of its measurements or its statistical parameters. In this work, a
3$\sigma$ clipping of the filtered data is applied to identify outliers.

\begin{figure}[t]
\begin{center}
\includegraphics[width=\linewidth]{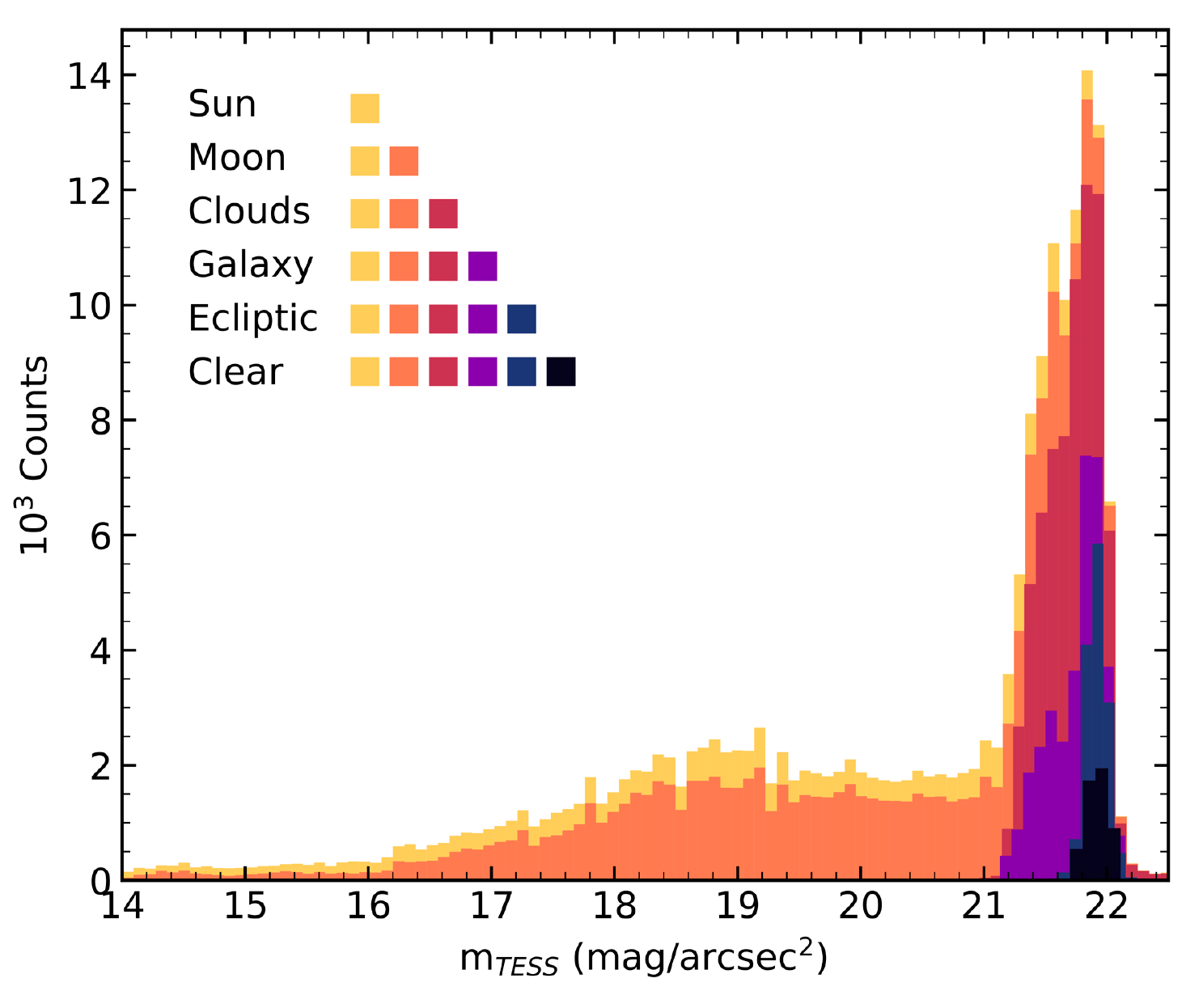}
\caption{Histogram of the data measured by the stars90 photometer between
2018 December 21 and 2019 November 18 at ORM (La Palma). The
components that affect the natural NSB have been filtered from the original
189,000 data obtained from $m_{\rm TESS}=14$ mag/arcsec$^2$ onward: sunlight (alt $< -18^\circ$, 31,000), moonlight (alt $< -5^\circ$, 85,000), clouds ($\sigma_{TESS} < 0.01$, 40,000), Galactic light ($|b| \geq 40^\circ$, 18,000) and zodiacal light (Equation (\ref{eq:ecl}), 10k), resulting in 5138 clear data (black) with P50$_{\rm TESS}$ = $21.94\pm0.04$, P99$_{\rm TESS}$ = $22.08\pm0.04$ and $\sigma_{\rm TESS} = 0.08$ mag/arcsec$^2$.\label{fig:hist}}
\end{center}
\end{figure}

\section{Natural Night Sky Brightness} \label{sec:naturalsky}
\begin{deluxetable*}{ccccccc}
\tablecaption{Percentile 50 and 99 of Zenithal NSB for ASTMON and TESS Measurements at ORM and OT during 2019--2020.
\label{tab:naturalsky}}
\tablewidth{0pt}
\tablehead{
\colhead{Band} & \colhead{P50 (ORM)} & \colhead{P99 (ORM)} & \colhead{Data (ORM)} & \colhead{P50 (OT)} & \colhead{P99 (OT)} & \colhead{Data (OT)} \\
\colhead{} & \colhead{(mag/arcsec$^2$)} & \colhead{(mag/arcsec$^2$)} & raw (filtered)& \colhead{(mag/arcsec$^2$)} & \colhead{(mag/arcsec$^2$)} & raw (filtered)
}
\startdata
$B$	&	$22.36 \pm 0.03$	&	$22.62 \pm 0.03$ & 4224 (181) & $22.05 \pm 0.03$	&	$22.32 \pm 0.03$ & 2236 (59)	\\
$V$	&	$21.66 \pm 0.02$	&	$21.89 \pm 0.02$ & 4217 (173) & $21.22 \pm 0.03$	&	$21.51 \pm 0.02$ & 2227 (91)	\\
$R$	&	$20.92 \pm 0.01$	&	$21.13 \pm 0.01$ & 4193 (171) & $20.72 \pm 0.01$	&	$21.09 \pm 0.01$ & 2078 (88)	\\
$m_{\rm TESS}$ 	&	$21.94 \pm 0.04$	&	$22.08 \pm 0.04$ & 179796 (5140) & 	$21.34 \pm 0.04$	&	$21.53 \pm 0.04$ & 430,993 (8101)	\\
\enddata
\tablecomments{Only filtered data (see section \ref{sec:analysis} for details) has been used to calculate final magnitude values.}
\end{deluxetable*}
When determining the natural sky brightness of different
observing sites and comparing studies, it is important to take
into account the definition of the natural NSB used and the
different filters or corrections applied to the raw data. With
regard to the former, it is essential to reach a quantitative
consensus on what is considered natural NSB. For instance, the
most common option is to take the arithmetic mean of all-sky
brightness measurements taken over several nights (see, for
example, \cite{Walker1988, Pilachowski1989, Mattila1996, Krisciunas1997,  Sanchez2007, Pedani2014, PlauchuFrayn2017, Posch2018}). Other authors
also include the minimum and maximum values as a reference
for natural NSB take into account the seasonal variations
produced, for example, by solar activity \citep{Duriscoe2007, Patat2008}. In this direction, \cite{Leinert1998} considered
the minimum/maximum values as the average of the three
smallest/largest sky brightness values (nightly averages) given
for each site. We consider that the natural NSB of a site should
be understood from a probabilistic point of view, quantifying
the frequency with which a given sky brightness value occurs
during the dark period--after filtering the data from the
components described in Section \ref{sec:analysis}--using percentiles, as done by \cite{Benn1998}, \cite{Kornilov2016}, \cite{Yang2017}, and \cite{Aube2020}. In this work, we have taken the P50 and P99 percentiles of the filtered data set as indicators of
the median and maximum darkness, as shown in Table \ref{table-TESS} together with the standard deviation. In total, 44 photometers
are included, after the constraint of having at least 1000 filtered
data and a percentage of clear night time greater than 30\% is
imposed.\\

The best value for the natural NSB (i.e., the pristine night
sky) for the Johnson--Cousins \textit{BVR} photometric system \citep{Bessell1990}, of the moonless sky on a clear night at high
ecliptic and Galactic latitudes and during solar minimum is
typically 22.7 mag/arcsec$^2$ in \textit{B}, 21.9 mag/arcsec$^2$ in \textit{V}, and
21.1 mag/arcsec$^2$ in \textit{R} \citep{Hanel2018}. For the SQM, the value of 22 mag/arcsec$^2$ is taken as a reference \citep{Falchi2016}. For all the photometers, the percentiles have been
calculated from the distribution resulting from applying the
filters explained in Section \ref{sec:analysis}. The result of this analysis is
shown in Figure \ref{fig:hist} for stars90 at ORM. Taking P50 as the
reference of the natural NSB, this is the darkest of all the sites
analyzed in this work (P50 = $21.94\pm0.04$ mag/arcsec$^2$), and
it is assumed that its NSB is, in the best atmospheric
conditions, very close to the natural one.\\

The natural NSB at ORM was measured on 427 CCD
images taken with the Isaac Newton and Jacobus Kapteyn
Telescopes on 63 photometric nights between 1987 and 1996
by \cite{Benn1998} (hereafter). The sky brightness at high elevation (elev.\ $>45^\circ$), high Galactic latitude ($|\beta| > 10^\circ$), high ecliptic latitude ($|\beta| > 40^\circ$), solar minimum (1994--1996) and moonless nights was $B=22.70 \pm 0.03$ mag/arcsec$^2$, $V=22.00 \pm 0.03$ mag/arcsec$^2$, $R=21.00 \pm 0.03$ mag/arcsec$^2$. To make comparisons with these standard
bands, we have also used the measurements taken by the
ASTMON \citep{Aceituno2011} located at ORM, contemporaneously
with the clear data from the stars90 photometer.
Table \ref{tab:naturalsky} shows the P50/P99 zenithal NSB measurements using
both devices. The new results are consistent with the BE98
measurements in \textit{B}. In the \textit{R} band, the P99 measurement is
darker than the BE98 estimate ($+0.13$ mag/arcsec$^2$), while $V$ is slightly brighter ($-0.11$ mag/arcsec$^2$). The ORM NSB
decreases in the \textit{R} band since the BE98 result is probably due to
the significant change in lighting systems on the island of La
Palma. From 1987 to 1996, a large number of low-pressure
sodium lamps emitted only in the \textit{R} band. The effect is the
opposite in the \textit{V} band, which is influenced by new LED
lighting systems \citep{Aube2020}. To summarize, \textit{BVR}
measurements by ASTMON and mTESS measurements by
TESS at ORM may be considered unpolluted, as artificial light
adds only approximately 2\% to the natural NSB. This is a
result similar to that of the Hawaiian Observatories \citep{Wainscoat2018}. On the other hand, night skies at OT are moderately
polluted ($-$0.31 in \textit{B}, $-$0.44 in \textit{V}, $-$0.2 in \textit{R} and $-$0.6 in $m_{\rm TESS}$, mag/arcsec$^2$ units), which is consistent with the
exhaustive analysis conducted by \cite{Aube2020}. In the
case of the San Pedro Martir Observatory (Mexico), measurements
made by stars213 during 2018–2020 show a P50$_{\rm TESS}$ of $21.77 \pm 0.04$ mag/arcsec$^2$, which is also consistent with the average $m_{\rm SQM} = 21.72$ mag/arcsec$^2$ measured in 2016 by \cite{PlauchuFrayn2017}.\\

\begin{figure*}[t]
\begin{center}
\includegraphics[width=0.9\linewidth]{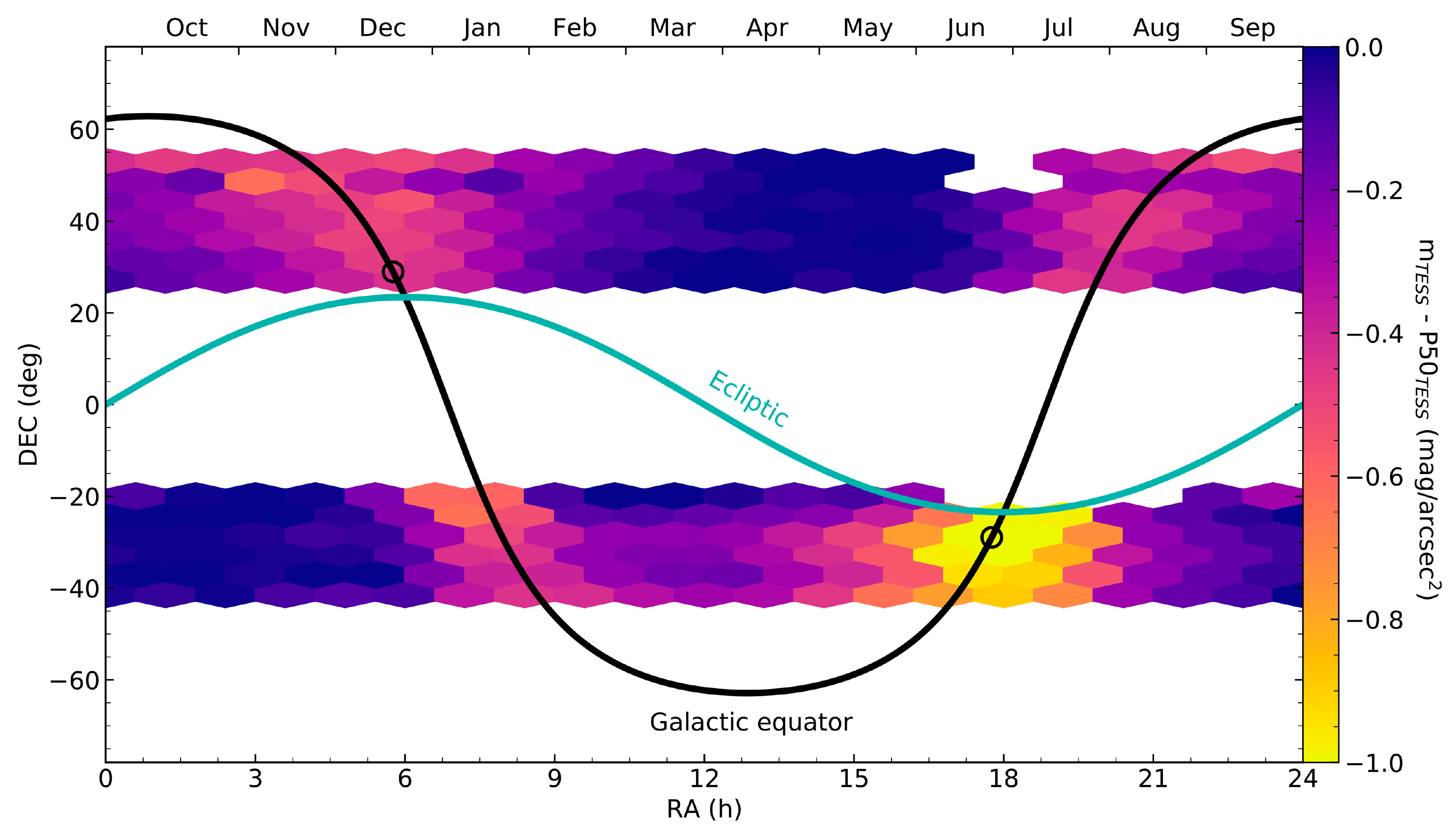}
\caption{NSB difference for a given celestial position (R.A., decl.). The upper axis includes the reference months, corresponding to the R.A. of the zenith at the
midpoint of the night, i.e., the sidereal time when the Sun is at lower culmination. The Galactic equator (black) and the ecliptic (green) are also included. The black
rings around ($-$18$^{\rm h}$, -30$^\circ$) and (6$^{\rm h}$, 30$^\circ$) mark the Galactic center and anticenter, respectively.\label{fig:radec}}
\end{center}
\end{figure*}

For a given location, there are nights of the year when
zenithal brightness measurements are inevitably influenced by
the different astronomical components. Figure \ref{fig:radec} shows the
difference in sky brightness for different positions ($\alpha$, $\delta$) in the
sky. Since the zenith decl. is the same as the geographical
latitude of the site, each device completes a horizontal line
throughout the year. All clear data have been averaged in bins
of approximately 1$^{\rm h}$ in R.A. and $5^\circ$ in decl. On the upper axis,
the months have been placed as a reference, taking for this
purpose the R.A. of the zenith in the middle of the night, i.e.,
the sidereal time when the Sun is at lower culmination. It is
noted that the best time to measure the zenithal natural NSB is
during local spring, as the available time is longer. In the earlier
months, measurements can be taken at the end of the night
(higher R.A.) while the opposite applies in the later months. At
equatorial latitudes, the presence of the ecliptic at the zenith
makes it impossible to make these measurements, which will
always be affected by zodiacal light. Furthermore, around the
solstices the Galaxy will be present, especially in the southern
winter, where the Galactic center will be very high at midnight.

\section{Discussion. Temporal variations} \label{sec:discu}
Many factors that can contribute to the determination of the
natural darkness of a particular site have been ignored in the
method presented here. For instance, near-desert locations (e.g.,
Namibia, Canary Islands, and Israel) could be affected by the
circulation of dust in the atmosphere \citep{Tomasz2020}. The
presence of fires (e.g., in Australia) can also affect NSB
measurements, as well as aerosols from anthropogenic activity
in nearby urban areas (e.g., in Extremadura, Balearic Islands,
and Arizona). ALAN may determine, of course, the median sky
brightness and could also hide some of the variations presented
(see \cite{Puschnig2020}). Nevertheless, the use of differential
photometry and the combination of many photometers placed
in dark locations make it possible to highlight emergent
properties by allowing a statistical characterization of the
natural NSB. As a matter of fact, all the sites present a standard
deviation very close to 0.1 mag/arcsec$^2$ (see Table \ref{table-TESS}), which
may indicate that there is a main mechanism driving the
variability of NSB measurements in very dark sites.\\

\begin{figure*}[t]
\begin{center}
\includegraphics[width=0.8\linewidth]{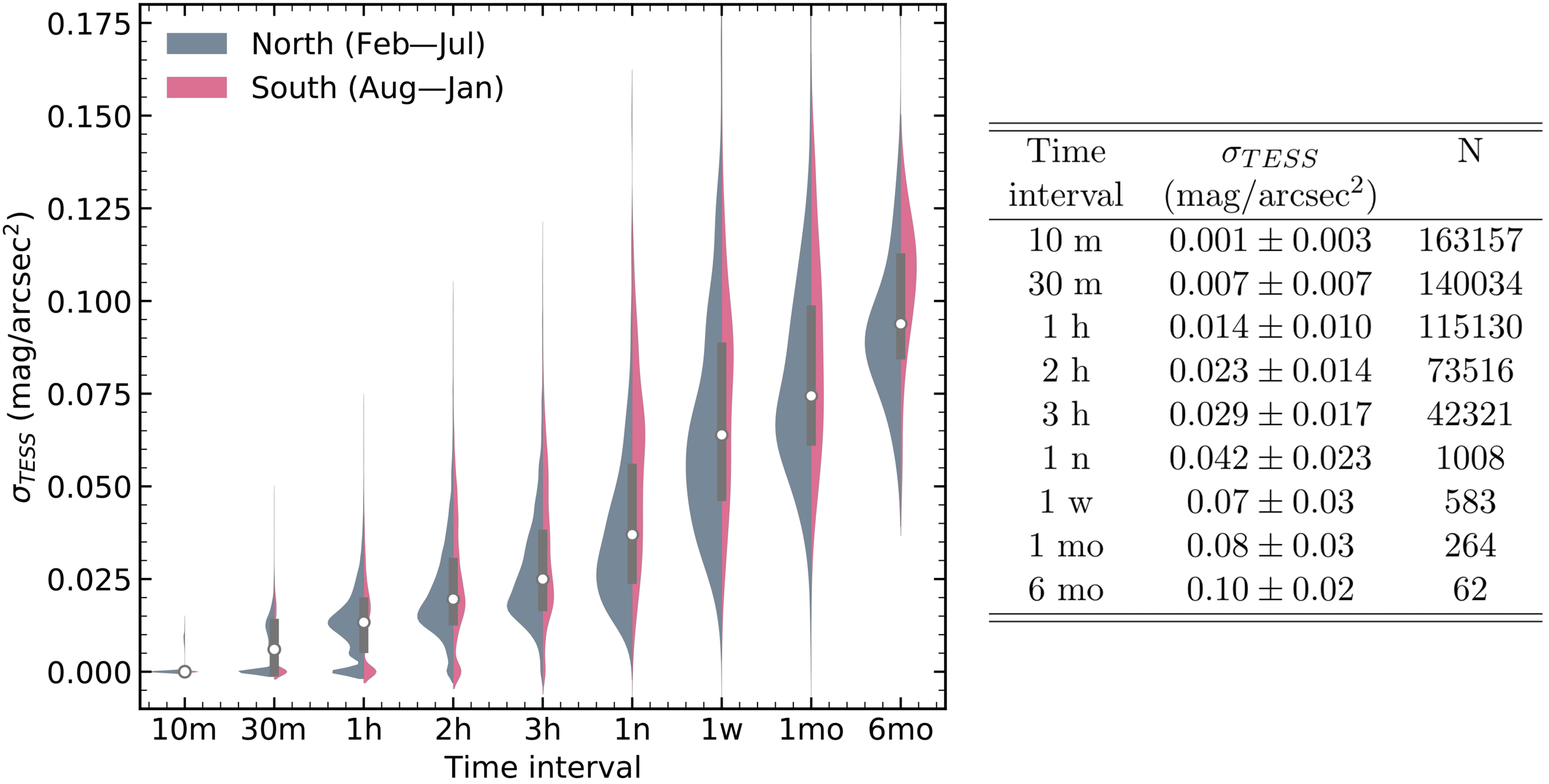}
\caption{Distributions of the standard deviation $\sigma_{\rm TESS}$ calculated from individual photometer measurements grouped in different time intervals: minutes (m), hours
(h), nights (n), weeks (w), and months (mo). The northern and southern hemispheres, corresponding to different periods of the year, have been divided to check that
there is no clear difference between them. Over the histograms, the mean (white dot) and the standard deviation (black box) of the distributions have been added. For
the sake of clarity, these values are shown in the supplementary table together with the total number of samples, N, in each interval.\label{tab:violin}}
\end{center}
\end{figure*}

The timescale has been investigated in order to study the
possible origin of this variability. Figure \ref{tab:violin} shows the
distribution of the standard deviation grouping of the data for
each photometer over different time intervals, from minutes to
months. Both hemispheres have been separated to check that
there is no apparent bias between them. If the total variability
of 0.1 mag/arcsec$^2$ were of an instrumental nature (some
undetected instrumental error of the device), it would be
expected to be reached at intervals of tens of minutes.
However, at intervals of 1 hr the median $\sigma_{\rm TESS}$ is 0.014 mag/arcsec$^2$ and at the minute timescale it is below 0.004 mag/arcsec$^2$, compatible with the instrumental error obtained in
laboratory tests (see \ref{TESSinserror}). The intranight variability,
with a median standard deviation of less than 0.04 mag/arcsec$^2$ , cannot explain the average 0.1 mag/arcsec$^2$ , but its
nature deserves to be explained in more detail in the following
sections.

\subsection{Dusk enhancement. Walker effect} \label{subsec:walker}
\begin{figure}[t]
\begin{center}
\includegraphics[width=\linewidth]{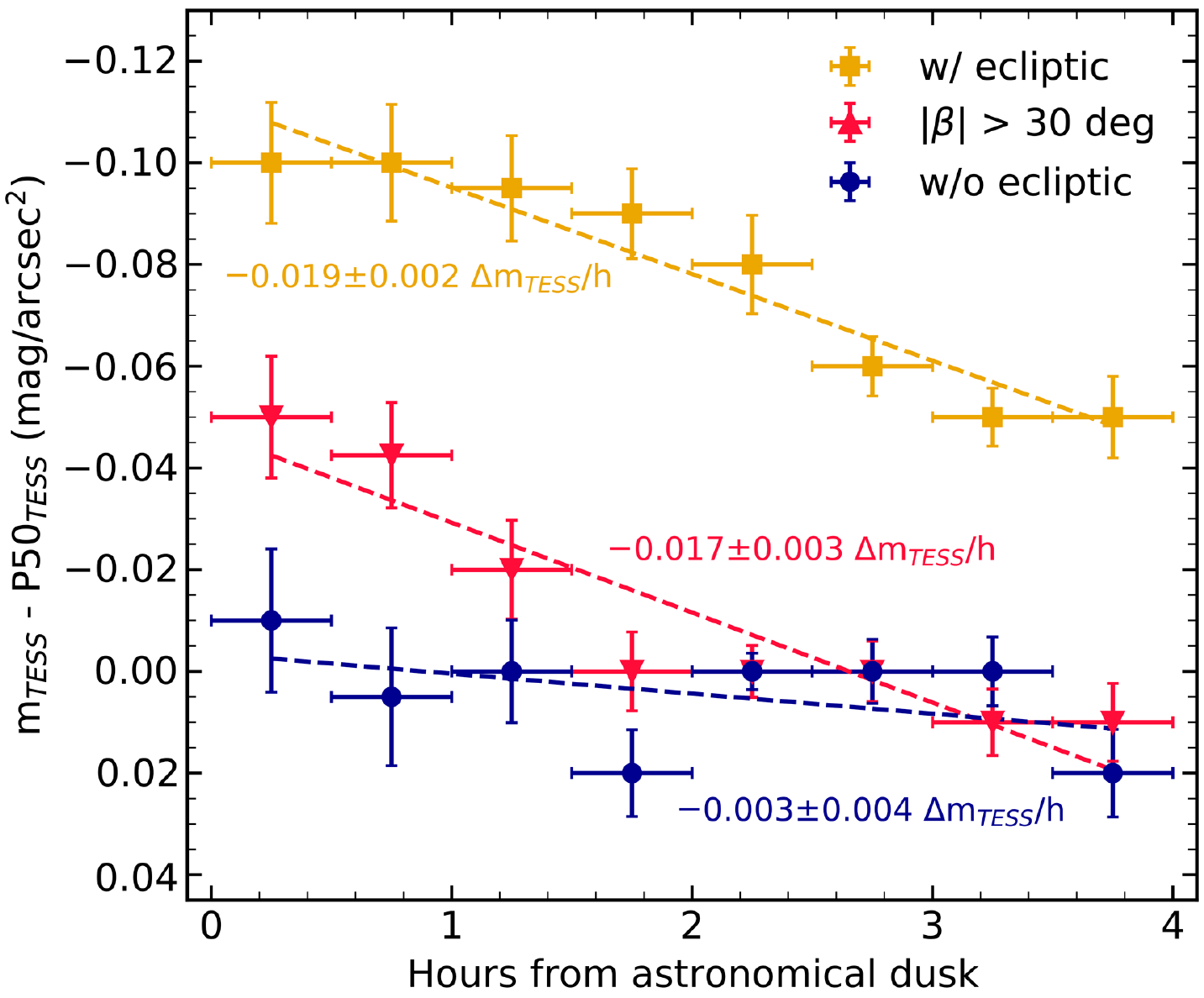}
\caption{NSB difference vs. time elapsed from the end of astronomical
twilight at the beginning of the night for data fully filtered with the method
proposed in this work (blue), for filtering of zodiacal light from absolute
elliptical latitudes higher than 30$^\circ$ (red), and for no filtering at all (yellow). Fit
lines and their slopes are included.\label{fig:walker}}
\end{center}
\end{figure}

One option for this intranight variability is the possible post-twilight
effects first identified by \cite{Walker1988}. Using
measurements made in the \textit{B} and \textit{V} filters from San Benito
Mountain during the period 1976--87, he found an exponential
decrease of the NSB of up to 0.4 mag/arcsec$^2$ during the first 4
hr of the night, which he identified with a process of
recombination of ions excited during the day by extreme
ultraviolet solar radiation. Since then, many authors have
addressed this issue at different sites and times within the solar
cycle. \cite{Taylor2004} reported a decrease of 0.2, 0.3, 0.3,
and 0.4 mag/arcsec$^2$ throughout the night by observations in
the \textit{UBVR} filters, respectively, from 1999 to 2003 at
Mount Graham, although their result was later refuted by \cite{Pedani2009}, who found no trend in a twice-larger sample
taken at the same location. Measurements of zenith NSB in the
\textit{UBVRI} filters taken at Calar Alto by \cite{Leinert1995}, at La Palma by \cite{Benn1998}, at Cerro Paranal by \cite{Patat2003}, and at San Pedro Martir by \cite{PlauchuFrayn2017} also show no evidence of a systematic nightly NSB decrease
after astronomical twilight.\\

A much smoother trend of 0.03 mag/hr was found by \cite{Krisciunas1997} in observations in \textit{B} and \textit{V}, close to the $0.01 \pm 0.01$ and $0.02\pm0.01$ mag/hr, respectively, in the review of Walker's data conducted by \cite{Patat2003}. Both took
into account the solar activity cycles and introduced corrections
for the zodiacal light contribution from the \cite{LevasseurDumont1980} model. While the exponential decrease
found by Walker has already been largely ruled out--it is
probably associated with light pollution \citep{Garstang1997} or
solar cycle effects--this smooth trend has not. In Figure \ref{fig:walker} we
have plotted the variation of the NSB measurements as a
function of time from the end of astronomical twilight. In the
data completely filtered by the method explained in Section \ref{sec:analysis} (blue dots), no clear trend toward a darkening of the sky during
the first hours of the night is observed. The slope of the fitted
line is $0.003 \pm 0.004$ $m_{\rm TESS}$/hrcompatible with 0. The plot
also includes the data for completely unfiltered zodiacal light
(yellow squares) and that filtered at a constant absolute ecliptic
latitude of 30$^\circ$. Note that although the former shows, as
expected, a higher brightening than the latter, both have similar
trends of $0.019\pm0.002$ and $0.017\pm0.003$ $m_{\rm TESS}$/hr, which
are also compatible with those reported by \cite{Krisciunas1997} and \cite{Patat2003}. As shown in Figure \ref{fig:walker}, in spite of its
increasing separation from the Sun in the early night, the
contribution of zodiacal light is smaller, resulting in a gradual
darkening during the first hours, especially in the \textit{B}, \textit{V}, and \textit{R}
filters, where the zodiacal light spectrum reaches its peak.

\subsection{Short-time variations} \label{subsec:short-term}
On hourly timescales, the variability may be due to the
passage of brighter or fainter star fields through the FOV of the
photometer, or to variations in airglow or skyglow scattering
throughout the night. In the former case, a star field would take
less than 1.5 hr to cross the entire FOV, so no difference should
be seen between the distributions at 2 hr, 3 hr, and 1 night in
Figure \ref{tab:violin}, although a progressive increase in $\sigma_{\rm TESS}$ is observed.
Furthermore, the same pattern should be seen on two
consecutive nights with a 4 minute delay in the same
photometer, which has not been found in general. Some
authors have identified particular nights when short-period
variations were present both in broadband photometry \citep{Leinert1998, Pilachowski1989, Patat2003} and in specific
lines characteristic of airglow emissions \citep{Patat2008}. In this
work, we report the systematic observation of short-time
variations in NSB (tens of minutes and hours) on the vast majority of nights and regardless of the geographical position,
fraction of the night, and season of the year.\\

\begin{figure}[t]
\begin{center}
\includegraphics[width=\linewidth]{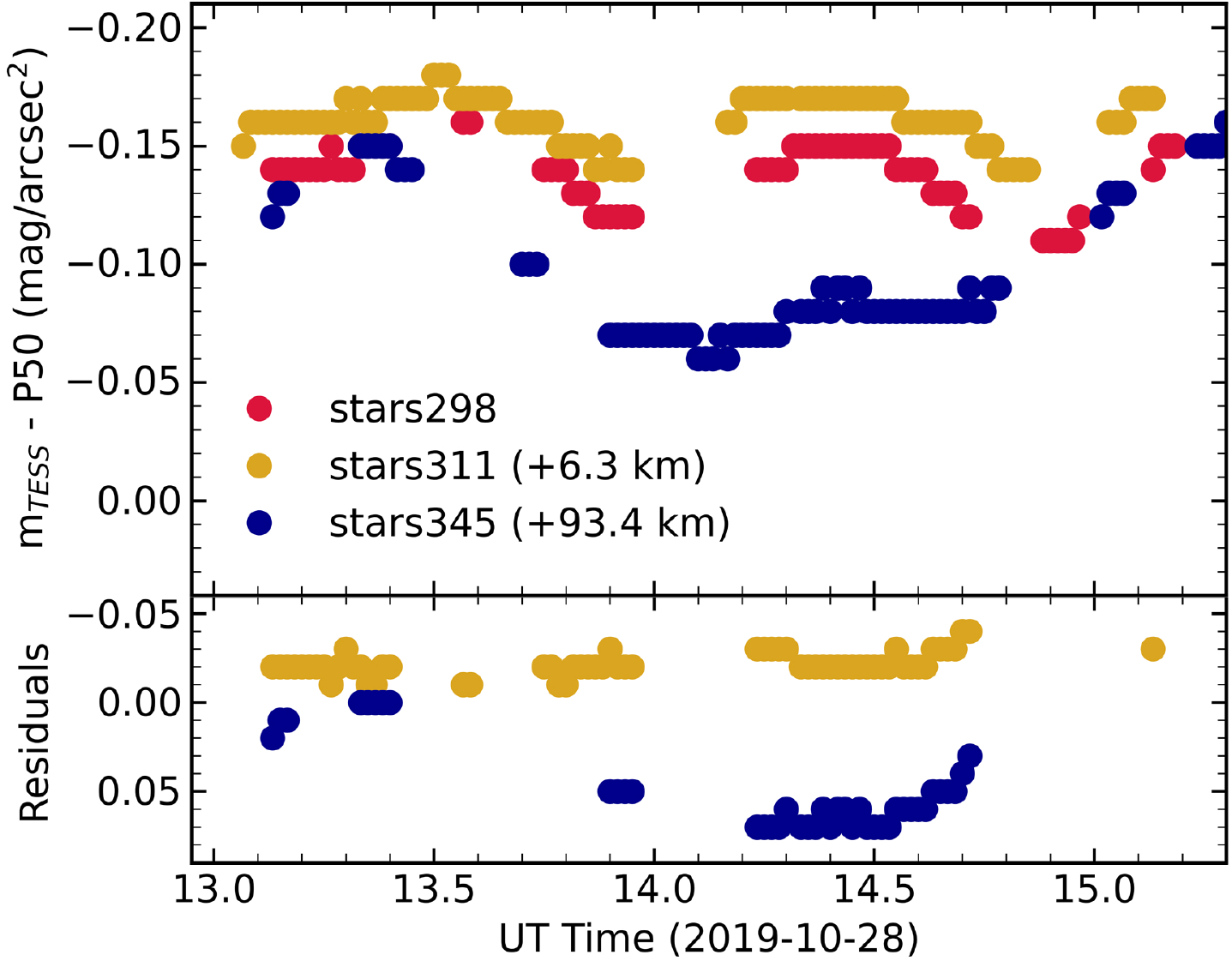}
\caption{Short-time variations of NSB on the night of 2019 October 28 for
three photometers located in Australia, two of them separated by 6.3 km and
the other by 93.4 km. The lower plot shows the differences between the
photometers stars311 (yellow) and stars345 (blue) with respect to
stars298 (red).\label{fig:waves}}
\end{center}
\end{figure}

\begin{figure*}[t]
\gridline{\fig{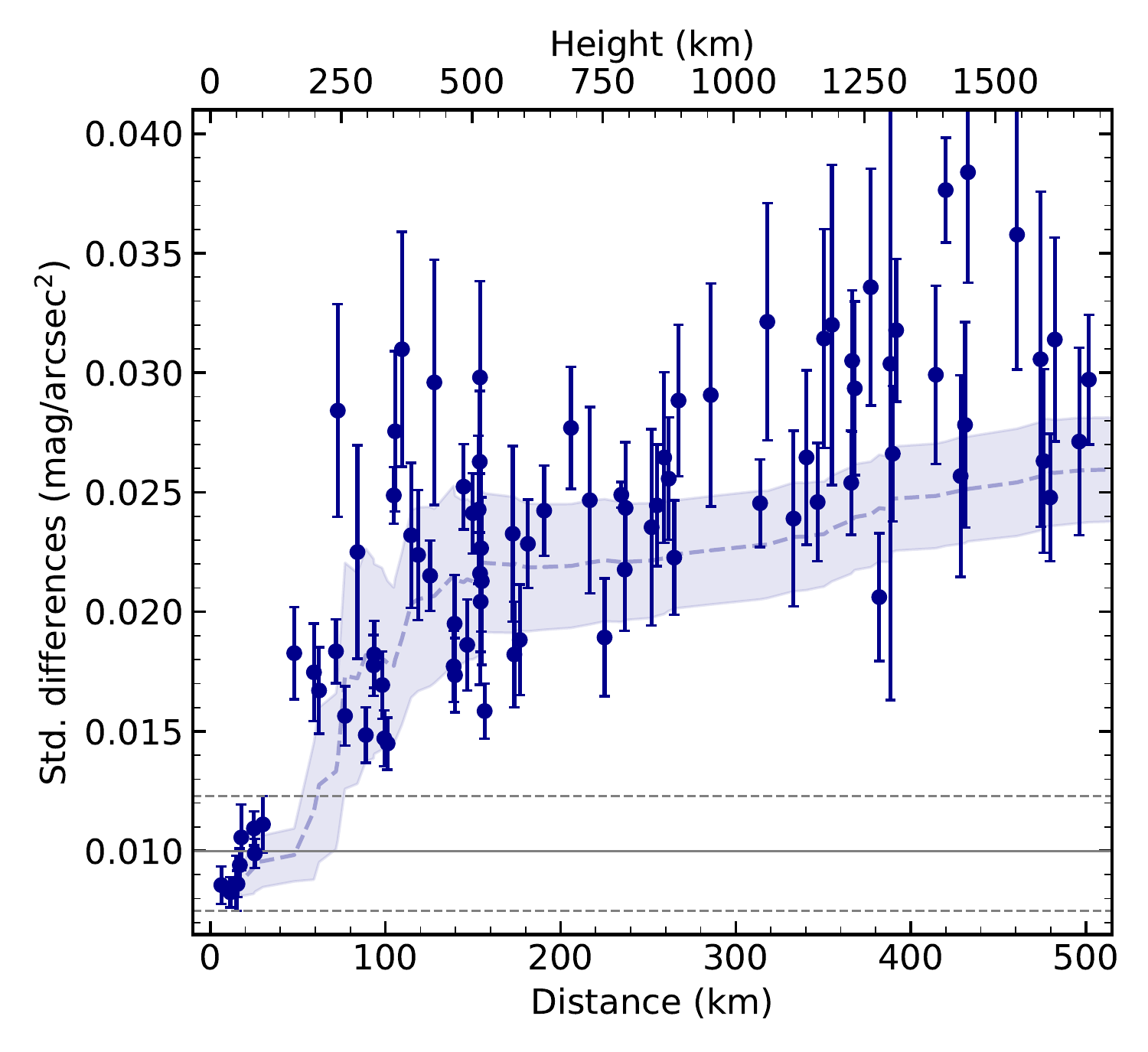}{0.48\linewidth}{}
\fig{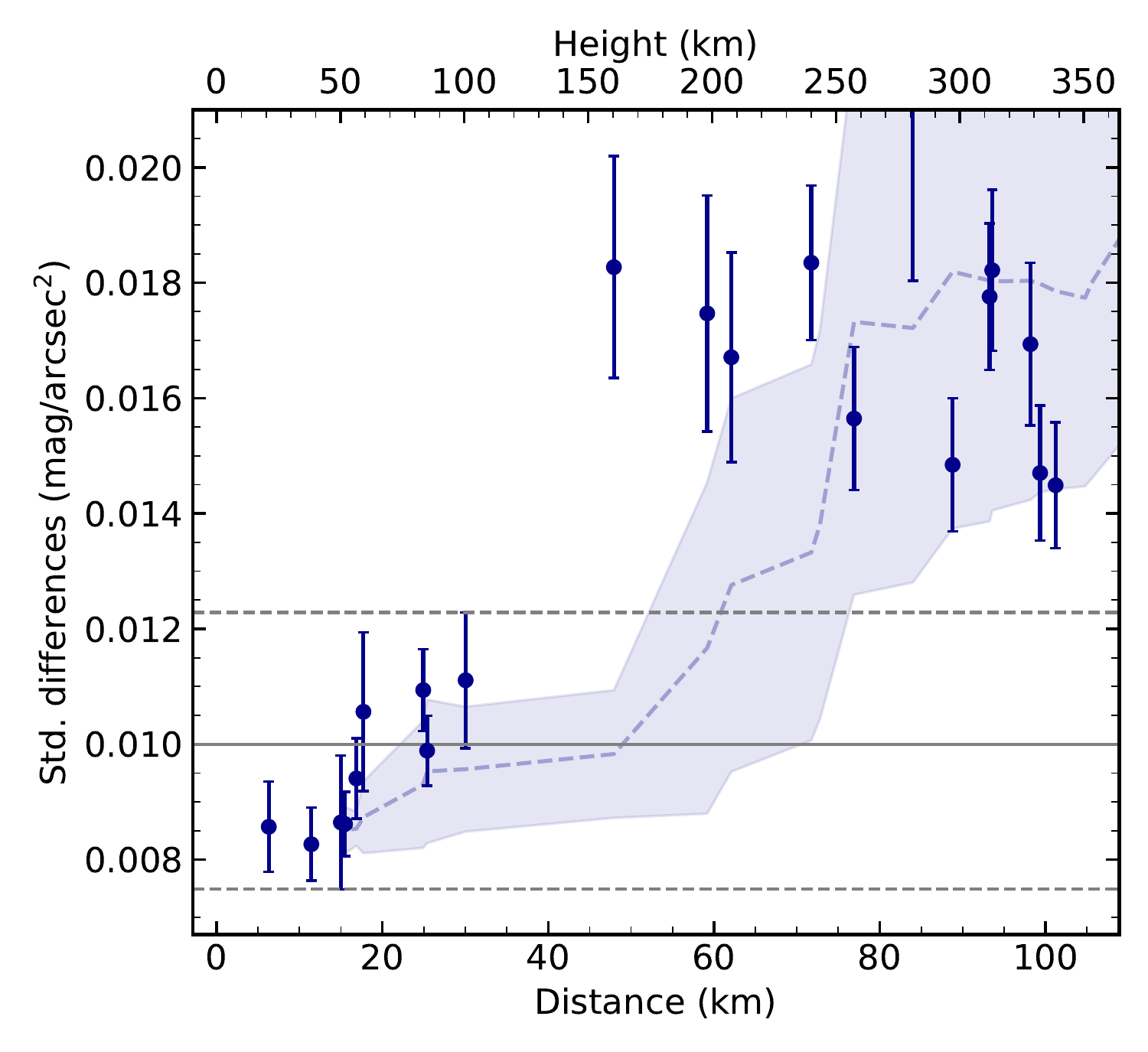}{0.48\linewidth}{}\vspace{-0.7cm}}
\caption{Standard deviation of the difference between simultaneous measurements of pairs of photometers vs. their separation. The geometric parallax height is
indicated on the upper axis. The moving average and the 3$\sigma$ limit are shown shaded and the coherence region is indicated by black horizontal lines. Points of closer
pairs (less than 100 km) are zoomed in on the right plot.\label{fig:airglow}}
\end{figure*}

An example of these variations is shown in Figure \ref{fig:waves}, which
includes data from three photometers located in Australia
between 13 and 15:30 UT on 2019 October 28. The stars298
(red) and stars311 (yellow) devices, located 6.3 km apart, show
very similar behavior, with a quasiperiodic oscillation of about
1 hr. The stars345 device (blue), located more than 90 km away
from them, shows similar behavior at the beginning and end of
the interval, but does not follow the brightening that occurs in
the other two at around 14.5 UT. This can be seen in the
difference between the stars345 and stars311 measurements
with respect to stars298, which is plotted below. The dispersion
between the photometers at 93.4 km is larger than the
dispersion between those at 6.3 km, whose NSB curves show
very similar. This is not a serendipitous result. With an FOV
(FWHM) of 17 17$^\circ$, two photometers close enough to each other
can see the same phenomenon occurring at a certain height
depending on the distance between them. In this case, we say
that the photometers are coherent and we can try to estimate the
height at which these short-time events occur by the geometric
parallax.\\

For this purpose, pairs of all available photometers have
been taken. Each night, we obtain the difference between
simultaneous measurements of both (at most 30 s apart) and
determine their standard deviation. The final value for each pair
and its uncertainty is obtained by averaging all the nights and calculating the standard error. The result for all the photometers
against their separation is shown in Figure \ref{fig:airglow}. The moving
average of all points (shaded) and the geometric parallax height
corresponding to each distance, obtained by simple trigonometry,
are included (effects of curvature, refraction, etc., have
been neglected as they do not affect this discussion). There are
two differentiated regions.\\

At long separations, beyond 150–200 km, the standard
deviation of the difference between their measurements shows
a stationary trend around 0.03 mag/arcsec$^2$ and very high
dispersion. This is a consequence of the loss of coherence
between the photometers, whose NSB measurements are not
correlated as they do not observe the same projected region
where the disturbances are occurring. Note that the value of
these standard deviations is compatible with that obtained for
the sky variation in 3 hr--1 night intervals in Figure \ref{tab:violin}, which
reinforces the idea that photometers far apart from each
other are not coherent and measure different NSB variations.
The differences between photometers separated by less than
30 km, on the other hand, show a very low dispersion, around
0.01 mag/arcsec$^2$ (see Figure \ref{fig:airglow}), which is compatible with
that obtained for the three adjacent photometers of OT
(Figure \ref{fig:TESS_calib}). The existence of coherence between photometers
located less than 30 km apart has a very relevant implication:
the phenomena observed simultaneously, which are the origin
of the short-time variations, are occurring at a height of at least
100 km. If these variations were due to tropospheric events
such as clouds, aerosols, turbulence, etc., no coherence would
be observed between photometers separated by more than 3
km. From this discussion we therefore conclude that the shortterm
NSB variations observed with the photometers are
produced by luminous phenomena that take place above the
mesosphere, i.e., airglow.

\subsection{Seasonal variations} \label{subsec:sasonal}
\begin{figure}[t]
\begin{center}
\includegraphics[width=\linewidth]{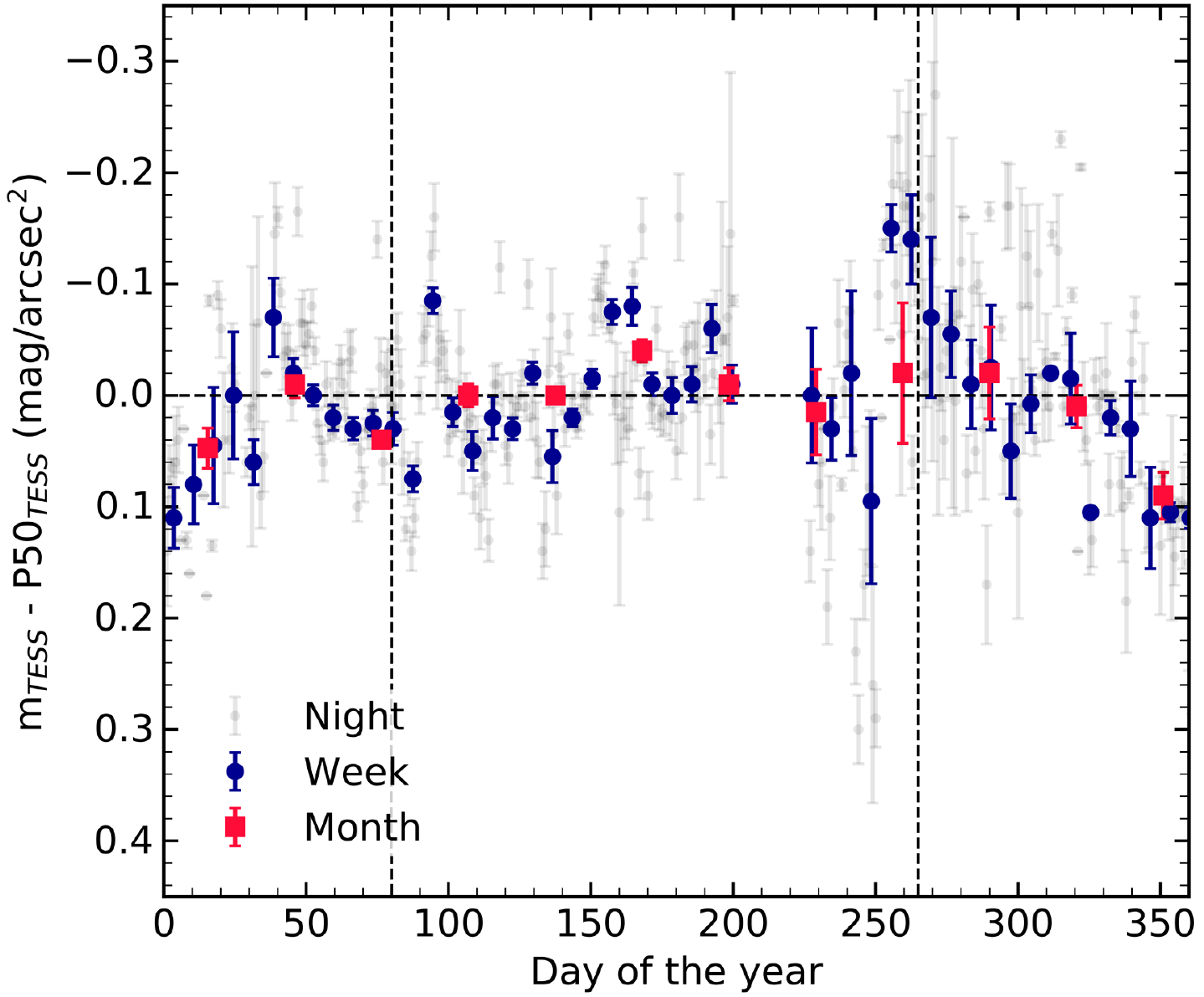}
\caption{NSB difference vs. day of the year since January 1 with data
averaged over night (gray), week (blue), and month (red) intervals. Equinoxes
are indicated by vertical dashed lines.\label{fig:seasonal}}
\end{center}
\end{figure}

Following the analysis of Figure \ref{tab:violin}, the total standard
deviation of 0.1 mag/arcsec$^2$ only seems to be reached on
monthly or annual scales. Several reasons could explain this.
One such reason is the dependence of the natural NSB on the
season of the year, which was identified by \cite{Patat2008} in both
\textit{BVRI} filters and airglow lines. To test this, in Figure \ref{fig:seasonal} we
have plotted our measurements with respect to the day of the
year, grouping them into nights, weeks, and months. A
semiannual trend similar to that observed by Patat in the R
and V filters seems to be suggested, with a brightening near the
equinoxes and a darkening near the solstices of amplitude
0.1 mag/arcsec$^2$, especially remarkable in the months of
December and January. In these regions of the spectrum, the
airglow lines of [OI] 5577 and [OI] 6300 (see Figure \ref{fig:spectra}) also show this variability (Figures 9 and 10 in \cite{Patat2008}). Thus, the
dispersion of 0.1 mag/arcsec$^2$ observed in all the photometers
could be due to variability in the airglow intensity on scales of
weeks or months. Its nature is still under study.\\

There may be other reasons for this yearly variability. The
possibility that the TESS photometers suffer from aging effects
is on the table. The only systematic study of this with SQM $+$
window devices is the recent one by \cite{Puschnig2021}, which suggests that a darkening of up to 0.053 m$_{\rm SQM}$ arcsec$^{-2}$ yr$^{-1}$ can occur at midlatitudes. However, we do not observe
evidence of this behavior in the continuous measurements of
the TESS photometers or any bias in the distributions of
photometers manufactured in different years. Further work
should be done on this matter.

The fact that, over long timescales, it is possible to detect an
overall increase in natural brightness as a consequence of the
increase in light pollution, quantified at 2.2\% by \cite{Kyba2017}, cannot be ignored either and requires a more detailed
examination over a longer period.\\

On the other hand, it is known that solar activity can affect
NSB by establishing peak-to-peak differences of up to 0.5 mag
in \textit{B} and \textit{V} during a solar cycle \citep{Walker1988, Leinert1995, Mattila1996, Krisciunas1997, Benn1998, Patat2008, PlauchuFrayn2017}. One of the most
widely used proxies to determine the relation between solar
activity and NSB measurements is the solar flux at 10.7 cm.
The correlation has been checked by taking different time
delays and no conclusive trend has been found. This
does not mean that it cannot be detected with broadband
photometers, but it is a consequence of the quiescence of the
Sun at minimum, when the range of F10.7 cm intensities is too
low to check these relations. More measurements will be
accumulated in the coming years to try to verify this
relationship.

\section{Conclusions} \label{sec:conclu}
This article has attempted to study natural NSB using data
from dozens of TESS photometers at more than 40 sites with
almost pristine skies during the solar minimum between Solar
Cycles 24 and 25. The main conclusions of this work may be
summarized as follows:\vspace{-0.2cm}
\begin{itemize}
\setlength\itemsep{-0.3em}
    \item[--] The first comprehensive reference method to measure the
natural NSB with low-cost broadband photometers has
been defined.
    \item[--] Differential photometry allows very high accuracies to be
achieved by combining different broadband photometers
in search of emerging properties.
    \item[--] The usefulness of the ROC curve has been demonstrated,
together with that of the standard deviation method, for
determining the most appropriate threshold for cloud
detection and its uncertainties.
    \item[--] Moonlight can contribute to the NSB measurement if its
elevation is higher than $-5^{\circ}$.
    \item[--] Galactic light can increase the NSB if it is measured at
an absolute latitude less than $40^{\circ}$ and approximately
homogeneously for all longitudes.
    \item[--] The heliocentric longitude has to be taken into account
in dealing with zodiacal light, and the possibility of
detecting the Gegenschein with broadband photometers
has been demonstrated for the first time.
    \item[--] The natural NSB has been calculated from the percentiles
for 44 different photometers, the ORM ones being the
darkest of all of them.
    \item[--] A standard deviation around 0.1mag/arcsec$^2$ is observed
in all the photometers, owing to NSB variations on scales
up to months.
    \item[--] The so-called Walker effect has been reviewed and no
evidence has been found for systematic darkening during the first hours of the night. This effect is compatible with
an insufficient subtraction of zodiacal light.
    \item[--] We report the systematic observation of short-time
variations in NSB on the vast majority of nights that
are related to airglow events forming above the
mesosphere.
    \item[--] There are indications of semiannual oscillations compatible
with what has been previously observed in airglow
emission features by other authors.
\end{itemize}

Future work should address possible aging effects and
relations of the natural NSB with solar and geomagnetic
phenomena. Distinguishing between the main components of
the airglow observed with TESS photometers and describing
their structure or dynamics requires, on the one hand, devices
with a narrower spectral range centered on the main emission
features and, on the other hand, better sampling of the transition
region between coherent and noncoherent photometers (following
the definition given in this work). The installation of
photometer networks with high temporal resolution separated
by tens of kilometers in very dark locations is highly desirable. \\

EELabs is a project funded by the European Union through
INTERREG V-A MAC 2014–2020. STARS4ALL was a
project funded by the European Union through H2020-ICT-
2015-688135. We thank the referee for constructive suggestions.
We thank Javier Díaz Castro and Federico de la Paz
(Oficina Técnica de Protección del Cielo, IAC) for maintaining
the ASTMON devices and Julio A. Castro Almazán for
processing the ASTMON data. The AMOS Canary Islands
facility of the Department of Astronomy, Earth Physics, and
Meteorology of Comenius University (Bratislava, Slovakia)
provided us with all-sky meteor images for cloud coverage
analysis validation. The High Energy Stereoscopic System
Observatory (H.E.S.S., Namibia), Hakos Astrofarm (Namibia),
and Extremadura Clear Skies are thanked for hosting TESS
photometer networks.

\bibliography{tess}{}
\bibliographystyle{aasjournal}

\begin{longrotatetable}
\begin{deluxetable*}{cccccccccccc}
\label{photlist}
\tablecaption{Coordinates, Elevations, and Locations of TESS Photometers, Ordered by NSB (P50).}
\label{table-TESS}
\tablewidth{700pt}
\tabletypesize{\scriptsize}
\tablehead{
\colhead{Name} & \colhead{Place} & \colhead{Latitude} & \colhead{Longitude} & \colhead{Alt.} & 
\colhead{Tot1} & \colhead{Tot2} & \colhead{P50} & \colhead{P99} & \colhead{$\sigma$} & \colhead{CNT} & \colhead{Dates} \\
\colhead{} & \colhead{} & \colhead{(deg)} & \colhead{(deg)} & \colhead{(m)} & 
\colhead{(raw)} & \colhead{(fil.)} & \colhead{$\pm 0.04$} & \colhead{$\pm 0.04$} & \colhead{} & \colhead{($\%$)} & \colhead{(yyyy-mm-dd)}
} 
\startdata
stars90 & ORM-IAC (La Palma,Spain) & 28.76391 & -17.89367 & 2152 & 189184 & 5138 & 21.94 & 22.08 & 0.08 & $77_{-5}^{+4}$ & 2018-12-21 2019-11-18\\
stars298 & EPSO (New South Wales,Australia) & -31.26010 & 149.22090 & 566 & 169564 & 3940 & 21.89 & 22.08 & 0.1 & $51_{-5}^{+3}$ & 2019-05-12 2020-07-11\\
stars343 & Wobblesock Obs. (New South Wales,Australia) & -31.22690 & 149.33450 & 574 & 189883 & 4313 & 21.87 & 22.07 & 0.1 & $55_{-5}^{+3}$ & 2019-06-07 2020-05-08\\
stars311 & Windana (New South Wales,Australia) & -31.27370 & 149.15670 & 617 & 197614 & 4946 & 21.86 & 22.06 & 0.11 & $59_{-5}^{+4}$ & 2019-05-04 2020-08-07\\
stars314 & Dragon's Rest (New South Wales,Australia) & -31.07900 & 149.37590 & 540 & 323160 & 5404 & 21.85 & 22.03 & 0.11 & $49_{-4}^{+3}$ & 2019-05-05 2020-10-15\\
stars342 & Mudgee Observatory (New South Wales,Australia) & -32.63310 & 149.49260 & 607 & 145635 & 3701 & 21.84 & 22.02 & 0.11 & $51_{-4}^{+3}$ & 2019-06-05 2020-04-08\\
stars345 & Yuma (New South Wales,Australia) & -30.84220 & 148.37300 & 168 & 226612 & 5775 & 21.82 & 22.04 & 0.11 & $60_{-5}^{+3}$ & 2019-06-14 2020-10-15\\
stars274 & Old Police Station (South Australia,Australia) & -30.81284 & 138.41243 & 241 & 300219 & 5815 & 21.8 & 22.12 & 0.13 & $56_{-4}^{+3}$ & 2019-03-26 2020-10-15\\
stars27 & Oukaïmeden Observatory (Marrakech-Safi,Morocco) & 31.20623 & -7.86648 & 2718 & 151939 & 1965 & 21.79 & 21.95 & 0.08 & $76_{-5}^{+4}$ & 2018-02-07 2019-12-06\\
stars213 & Obs. Astronómico Nacional (Baja California,Mexico) & 31.04350 & -115.46410 & 2797 & 336241 & 6525 & 21.77 & 22.04 & 0.11 & $74_{-5}^{+4}$ & 2018-10-06 2020-03-27\\
stars361 & Junabee Q. (Victoria,Australia) & -36.80537 & 144.67604 & 240 & 105178 & 2810 & 21.76 & 21.93 & 0.1 & $36_{-4}^{+3}$ & 2020-05-13 2020-10-15\\
stars62 & Centre d'Observació de l'Univers (Catalonia,Spain) & 42.02459 & 0.73480 & 804 & 525605 & 10397 & 21.76 & 21.98 & 0.09 & $40_{-4}^{+3}$ & 2018-02-07 2020-10-15\\
stars206 & Mountain Lodge, Etosha NP (Kunene,Namibia) & -19.25820 & 15.26240 & 1186 & 68105 & 1321 & 21.76 & 21.89 & 0.09 & $44_{-4}^{+3}$ & 2018-10-09 2019-03-14\\
stars270 & H.E.S.S. (Khomas,Namibia) & -23.27295 & 16.50283 & 1817 & 420186 & 3361 & 21.75 & 21.98 & 0.11 & $60_{-5}^{+3}$ & 2018-11-29 2020-10-15\\
stars202 & Casa Rural Cíjara (Extremadura,Spain) & 39.32287 & -4.95050 & 487 & 230867 & 6659 & 21.71 & 21.95 & 0.13 & $60_{-5}^{+3}$ & 2018-06-16 2020-07-07\\
stars230 & Long Dark Sky (Wellington,New Zealand) & -41.23592 & 175.43898 & 29 & 401605 & 5143 & 21.71 & 22.06 & 0.12 & $34_{-4}^{+3}$ & 2018-11-16 2020-10-15\\
stars18 & Javalambre (CEFCA) (Aragon,Spain) & 40.04180 & -1.01630 & 1942 & 248579 & 6067 & 21.7 & 21.93 & 0.09 & $38_{-4}^{+3}$ & 2018-02-07 2020-08-20\\
stars219 & Hakos Astrofarm (Khomas,Namibia) & -23.23636 & 16.36177 & 1824 & 261055 & 1876 & 21.7 & 21.9 & 0.09 & $60_{-5}^{+4}$ & 2018-10-02 2020-10-15\\
stars341 &  Frog Rock (New South Wales,Australia) & -32.45520 & 149.66290 & 502 & 266036 & 4702 & 21.66 & 21.9 & 0.11 & $46_{-5}^{+3}$ & 2019-06-01 2020-07-24\\
stars227 & Oracle (Arizona,USA) & 32.62140 & -110.74660 & 1307 & 143946 & 3331 & 21.65 & 21.8 & 0.07 & $63_{-5}^{+3}$ & 2019-03-27 2020-02-10\\
stars373 & Atutahi Observatory (Wellington,New Zealand) & -41.23617 & 175.48353 & 72 & 185930 & 1687 & 21.64 & 21.82 & 0.1 & $33_{-4}^{+3}$ & 2020-01-04 2020-10-15\\
stars289 & Astrocamp (Castile-La Mancha,Spain) & 38.16578 & -2.32689 & 1590 & 336219 & 7832 & 21.61 & 21.79 & 0.09 & $53_{-5}^{+3}$ & 2019-03-07 2020-10-15\\
stars349 & Consell Insular de Menorca (Balearic Islands,Spain) & 40.04912 & 4.05414 & 9 & 240723 & 3722 & 21.6 & 21.85 & 0.09 & $45_{-4}^{+3}$ & 2019-07-21 2020-09-13\\
stars232 & Parco Astronomico Lilio (Calabria,Italy) & 39.31450 & 16.75130 & 1166 & 422863 & 7467 & 21.58 & 21.8 & 0.09 & $48_{-5}^{+3}$ & 2018-10-14 2020-10-15\\
stars66 & Hospedería de Monfragüe (Extremadura,Spain) & 39.78100 & -6.01626 & 309 & 435827 & 9272 & 21.54 & 21.78 & 0.11 & $68_{-5}^{+4}$ & 2018-06-16 2020-10-15\\
stars292 & Esparragosa de Lares (Extremadura,Spain) & 38.94895 & -5.22763 & 456 & 213255 & 5531 & 21.54 & 21.88 & 0.18 & $52_{-4}^{+3}$ & 2019-03-15 2020-09-20\\
stars221 & Complejo Astronómico El Leoncito (SJ,Argentina) & -31.79861 & -69.29774 & 2463 & 269944 & 5397 & 21.52 & 21.82 & 0.18 & $50_{-4}^{+3}$ & 2018-11-08 2020-10-15\\
stars288 & Entre Encinas y Estrellas (Extremadura,Spain) & 38.21969 & -6.63176 & 493 & 328279 & 6930 & 21.51 & 21.75 & 0.11 & $61_{-5}^{+4}$ & 2019-01-10 2020-10-15\\
stars239 & Puimichel (Provence-Alpes-Côte d'Azur,France) & 43.97808 & 6.01788 & 697 & 152893 & 4826 & 21.5 & 21.69 & 0.09 & $60_{-5}^{+3}$ & 2018-12-17 2019-12-01\\
stars220 & Pierre Auger Observatory (Mza.,Argentina) & -35.29218 & -69.01251 & 1391 & 224102 & 3334 & 21.49 & 21.67 & 0.13 & $61_{-5}^{+4}$ & 2019-02-09 2020-07-01\\
stars296 & Casa Rural El Recuerdo (Extremadura,Spain) & 39.42861 & -5.78417 & 533 & 282058 & 4904 & 21.45 & 21.64 & 0.1 & $58_{-5}^{+3}$ & 2019-05-03 2020-10-15\\
stars291 & Alcazaba de Reina (Extremadura,Spain) & 38.18984 & -5.95709 & 813 & 272833 & 3936 & 21.45 & 21.66 & 0.08 & $61_{-5}^{+4}$ & 2019-03-15 2020-10-15\\
stars11 & El Torcal (Andalusia,Spain) & 36.95000 & -4.54000 & 1038 & 36804 & 1889 & 21.4 & 21.57 & 0.09 & $68_{-5}^{+3}$ & 2018-02-07 2018-07-05\\
stars211 &  OT-IAC (Tenerife,Spain) & 28.30033 & -16.51221 & 2384 & 477139 & 10824 & 21.38 & 21.64 & 0.12 & $78_{-5}^{+4}$ & 2018-02-07 2020-10-15\\
stars218 & La Roca de la Sierra (Extremadura,Spain) & 39.10380 & -6.66335 & 258 & 384440 & 9884 & 21.38 & 21.62 & 0.09 & $71_{-5}^{+4}$ & 2018-10-20 2020-10-15\\
stars246 & Obs. de Sierra Nevada (Andalusia,Spain) & 37.06417 & -3.38472 & 2802 & 211059 & 3350 & 21.38 & 21.61 & 0.09 & $51_{-4}^{+3}$ & 2019-06-18 2020-10-15\\
stars396 & Hotel Morvedra Nou (Balearic Islands,Spain) & 39.97366 & 3.90147 & 79 & 129536 & 3318 & 21.37 & 21.59 & 0.09 & $35_{-4}^{+3}$ & 2019-12-17 2020-10-15\\
stars271 & Puerto Villareal (Extremadura,Spain) & 38.74252 & -7.21002 & 146 & 96217 & 3287 & 21.34 & 21.46 & 0.07 & $85_{-6}^{+4}$ & 2018-12-04 2020-10-15\\
stars242 & Joint Ins. for VLBI ERIC (Drenthe,Netherlands) & 52.81282 & 6.39595 & 16 & 222049 & 5404 & 21.34 & 21.55 & 0.11 & $31_{-4}^{+3}$ & 2018-11-14 2020-07-25\\
stars8 & Guirguillano (Navarre,Spain) & 42.71167 & -1.86509 & 582 & 345776 & 4999 & 21.25 & 21.41 & 0.07 & $38_{-4}^{+3}$ & 2018-02-07 2020-10-15\\
stars257 & Isla Blanca (Balearic Islands,Spain) & 39.06986 & 1.40878 & 250 & 158267 & 2312 & 21.22 & 21.34 & 0.09 & $53_{-5}^{+3}$ & 2019-04-17 2020-07-08\\
stars201 & Finca La Cocosa (Extremadura,Spain) & 38.75545 & -6.98451 & 244 & 358402 & 9260 & 21.14 & 21.46 & 0.16 & $88_{-5}^{+4}$ & 2018-06-14 2020-09-19\\
stars401 & Obs. Jost (North Rhine-Westphalia,Germany) & 50.52234 & 6.52818 & 467 & 152524 & 5090 & 21.14 & 21.26 & 0.06 & $33_{-4}^{+3}$ & 2019-12-04 2020-09-14\\
stars347 & Wise Observatory (South District,Israel) & 30.59738 & 34.76226 & 863 & 261080 & 4961 & 21.07 & 21.33 & 0.13 & $90_{-6}^{+4}$ & 2019-08-21 2020-10-15\\
\enddata
\tablecomments{A total of 44 TESS devices are included in the present work comprising a raw data total of 11,099,432 individual measurements. After filtering (see Section \ref{sec:analysis}) the total number of clear data is reduced to 222,605.
All the photometers are still collecting data, but the present analysis is limited to the end date 2020 October 15. Exact dates are shown in the last column (Dates). P50 and P99 are the $m_{TESS}$ in units of magnitude
per square arcsecond. The values of $\sigma$ are the standard deviations of the magnitude distributions. The CNT column is an estimation of the percentage of clear night time on every photometer by means of the ROC curve
method explained in Section \ref{subsec:cloud}.}
\end{deluxetable*}
\end{longrotatetable}


\clearpage
\appendix
\label{sec:appendix}
\section{TESS photometric errors} \label{TESSerrors}
\begin{figure}
	\centering
	\begin{minipage}[t]{.48\columnwidth}
		\centering
		\includegraphics[width=\textwidth]{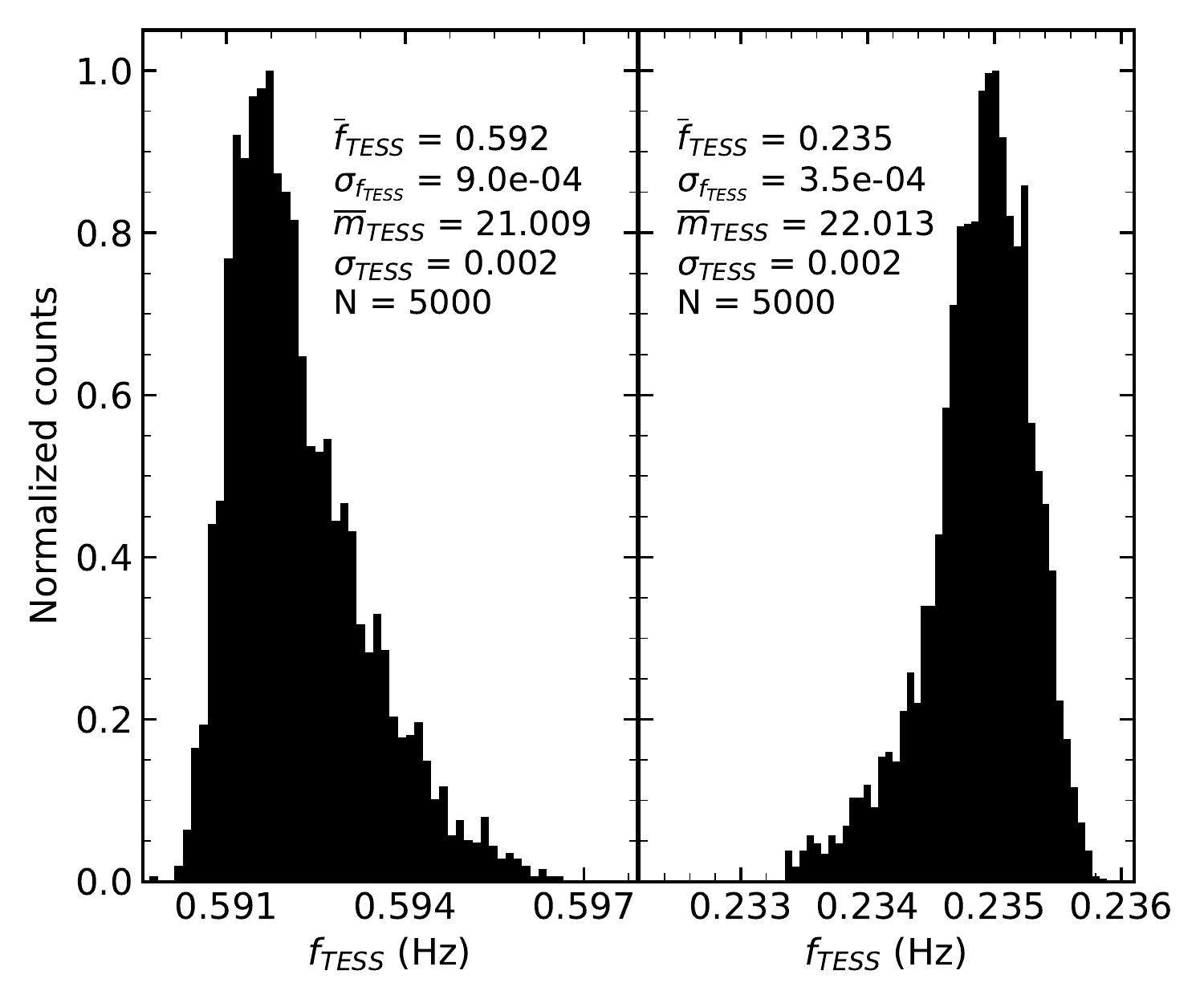}
		\caption{Distribution of 5000 laboratory measurements with light intensities
equivalent to $m_{\rm TESS}$ = 21 (left) and 22 (right) mag/arcsec$^2$. The standard
deviation of the distribution, in both cases $\sigma_{\rm TESS}$ = 0.002, is considered an
upper bound on the instrumental error.}
\label{fig:instr_TESS}
	\end{minipage}%
	\hfill
	\begin{minipage}[t]{.48\columnwidth}
		\centering
		\includegraphics[width=\textwidth]{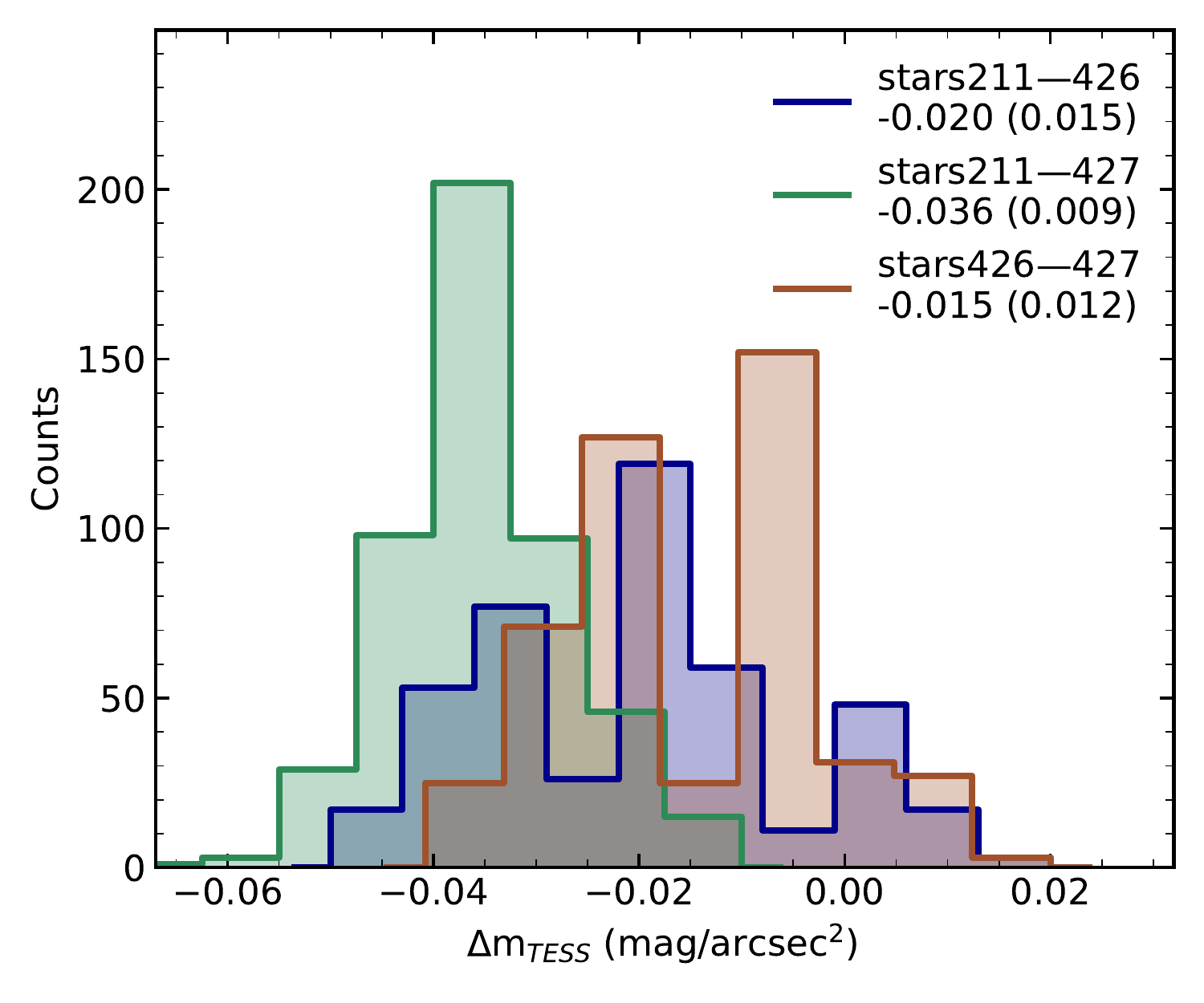}
		\caption{Measurement intercomparison between three TESS photometers--
stars426, stars427, and stars211--placed adjacent to one another at OT. The
histograms show the distribution of the difference in zenithal NSB
measurements. In total, about 300 measurements have been compared, whose
mean difference and (standard deviation) are shown in the plot.}
		\label{fig:TESS_calib}
	\end{minipage}
\end{figure}

Following standard astronomical practice, brightness is
usually expressed in units of magnitude per square second.
For the TESS photometric band it is defined as \citep{Bar2019}:
\begin{equation}
\label{magTESS}
    m_{\rm TESS} = ZP -2.5\log_{10}(f_{\rm TESS}-f_{\rm TESS}^{\rm D})
\end{equation}
where $ZP$ is the laboratory-defined zero-point of the photometer
using a reference light source and an absolute-calibrated
sensor inside an integrating sphere, and $f_{\rm TESS}$ and $f_{\rm TESS}^{\rm D}$ are the
signal and dark frequencies of the sensor, respectively, both
measured in hertz. At typical nighttime low operating
temperatures ($< 20 ^\circ$), the value of  $f_{\rm TESS}^{\rm D}$ turns out to be
negligible in most cases. The total uncertainty (uncertainty and
error are used indistinctly in this section), $\delta^2 m_{\rm TESS}$, of the
measurements of the TESS photometer for the output
frequency $f_{\rm TESS}$ is obtained and converted to device-specific
brightness units by propagating expression (\ref{magTESS}) as follows,
\begin{equation}
    \delta^2 m_{\rm TESS} =\delta_{ZP}^2 +
    ~\left(\frac{2.5\log_{10}(e)}{f_{\rm TESS}}~\delta_{f_{\rm TESS}} \right)^2
     = \rm Calibration ~Error^2 + Instrumental ~Error^2
\end{equation}
Considering a calibration and instrumental error of 0.044 and
0.002 mag/arcsec$^2$ , respectively (see next sections for details),
the final calculated error will be equal to 0.04404 mag/arcsec$^2$ . An uncertainty of 0.04 mag/arcsec$^2$ is adopted for the
magnitudes measured by the TESS photometers. A measurement
intercomparison of several TESS photometers at the same
site in dark-sky conditions is done to validate this value.
Figure \ref{fig:TESS_calib} shows the distribution of the difference in zenithal
NSB measurements between three TESS photometers--
stars426, stars427, and stars211--placed adjacent to one
another at OT during a period of 2 months (2020 June and July). The NSB statistical percentiles of the TESS photometers
are shown in Table \ref{tab:errorOT}. In both percentile cases, the final adopted
magnitude uncertainty is in good agreement with the experimental
data.\\

To study the nighttime evolution of NSB and possible
correlations with different atmospheric and astronomical
sources, differential photometry must be done to measure
magnitudes from two different TESS photometers. Considering
two photometer measurements taken at the same time t, and
neglecting dark currents, magnitudes are defined following
Equation (\ref{magTESS}) as
\begin{equation}
\begin{aligned}
m_{\rm TESS1}(t) =& ZP1 -2.5\log_{10}(f_{\rm TESS1}(t))\\
m_{\rm TESS2}(t) =& ZP2 -2.5\log_{10}(f_{\rm TESS2}(t))\label{magTESS1}
\end{aligned}
\end{equation}
and the result of differential photometry is
\begin{equation}
\label{difTESS}
    \Delta m_{\rm TESS}(t) = m_{\rm TESS2}(t) - m_{\rm TESS1}(t) = 2.5\log_{10}\left(\frac{f_{\rm TESS1}(t)}{f_{\rm TESS2}(t)}\right)+(ZP2-ZP1)
\end{equation}
Assuming $ZP1$ and $ZP2$ are constants (calibration error is
considered to be a systematic error), the total squared
uncertainty for differential photometry, $\delta^2 \Delta m_{\rm TESS}$, is obtained
and converted to device-specific brightness units by propagating
the previous expression:
\begin{equation}
    \delta^2 \Delta m_{\rm TESS} = (2.5\log_{10}(e))^2
    ~\left(
    ~\left( \frac{\delta_{f_{\rm TESS1}(t)}}{f_{\rm TESS1}(t)} \right)^2 +
    ~\left( \frac{\delta_{f_{\rm TESS2}(t)}}{f_{\rm TESS2}(t)} \right)^2 
    \right)
\end{equation}
Working with a NSB around 21 mag/arcsec$^2$, $f_{\rm TESS1}(t) \approx f_{\rm TESS2}(t)=f_{(m_{\rm TESS}=21)}=0.592$ Hz, therefore $\delta_{f_{\rm TESS1}(t)}\approx\delta_{f_{\rm TESS2}(t)}=\delta_{f_(m_{\rm TESS}=21)}=0.0009$ mag/arcsec$^2$ (see Figure \ref{fig:instr_TESS}) and final differential photometry error will be 
\begin{equation}
    \delta \Delta m_{\rm TESS} = 
    ~\left(
    \frac{(2.5\log_{10}(e))\sqrt{2}}{f_{(m_{\rm TESS}=21)}}
    \right) 
    ~\delta_{f_{(m_{\rm TESS}=21)}} = 0.0023~ mag/arcsec^2
\end{equation}

This result (that is, the analog for $m_{\rm TESS}$ = 22 mag/arcsec$^2$)
should be considered theoretical. When differential photometry
is applied to real data, the photometric uncertainty is around
0.02 mag/arcsec$^2$, as shown in Figure \ref{fig:TESS_calib} (standard deviation).
This error increase is caused by unknown and unpredictable
small changes in the environmental conditions and photometer
internal conditions.

\begin{deluxetable}{ccccc}
\tablecaption{Percentile 50 and 99 of Zenithal NSB for Three Identical TESS Photometers
Installed Adjacent to One Another at OT.
\label{tab:errorOT}}
\tablehead{
\colhead{Name} & \colhead{P50 (OT)} & \colhead{P99 (OT)} &  \colhead{Total} \\
\colhead{} & \colhead{(mag/arcsec$^2$)} & \colhead{(mag/arcsec$^2$)} & \colhead{raw (filtered)}
}
\startdata
stars426 & $21.26 \pm 0.04$	& $21.41 \pm 0.04$ & 32,148 (300)\\
stars427 & $21.28 \pm 0.04$ & $21.43 \pm 0.04$ & 22,350 (342)\\
stars211 & $21.24 \pm 0.04$	& $21.39 \pm 0.04$ & 32,903 (300)\\
\enddata
\tablecomments{Data were taken during 2020 June and July. Only filtered data (see Section \ref{sec:analysis} for details) have been used to calculate the percentiles.}
\end{deluxetable}

\subsection{TESS Calibration Error} \label{TESScalerror}
The TESS devices are calibrated by the manufacturer in the
Laboratory for Scientific Advanced Instrumentation (Laboratorio
de Instrumentación Científica Avanzada) of UCM
(Spain) using a photodiode and a TESS photometer unit, both
calibrated absolutely (see \cite{Bar2019} for the absolute
calibration procedure). For the first 39 TESS photometers
manufactured, the mean value of the measured zero-points
was ZP = 20.44 mag/arcsec$^2$ showing a dispersion of $\sigma$ =
0.044 mag/arcsec$^2$ (see Figure 5 of \cite{PosterZamorano2017}). This well-defined calibration method for the TESS photometer
allows for intercomparison among the measurements of several
TESS devices of different research teams.

\subsection{TESS Instrumental Error} \label{TESSinserror}
\begin{figure}[t]
\begin{center}
\includegraphics[width=0.75\linewidth]{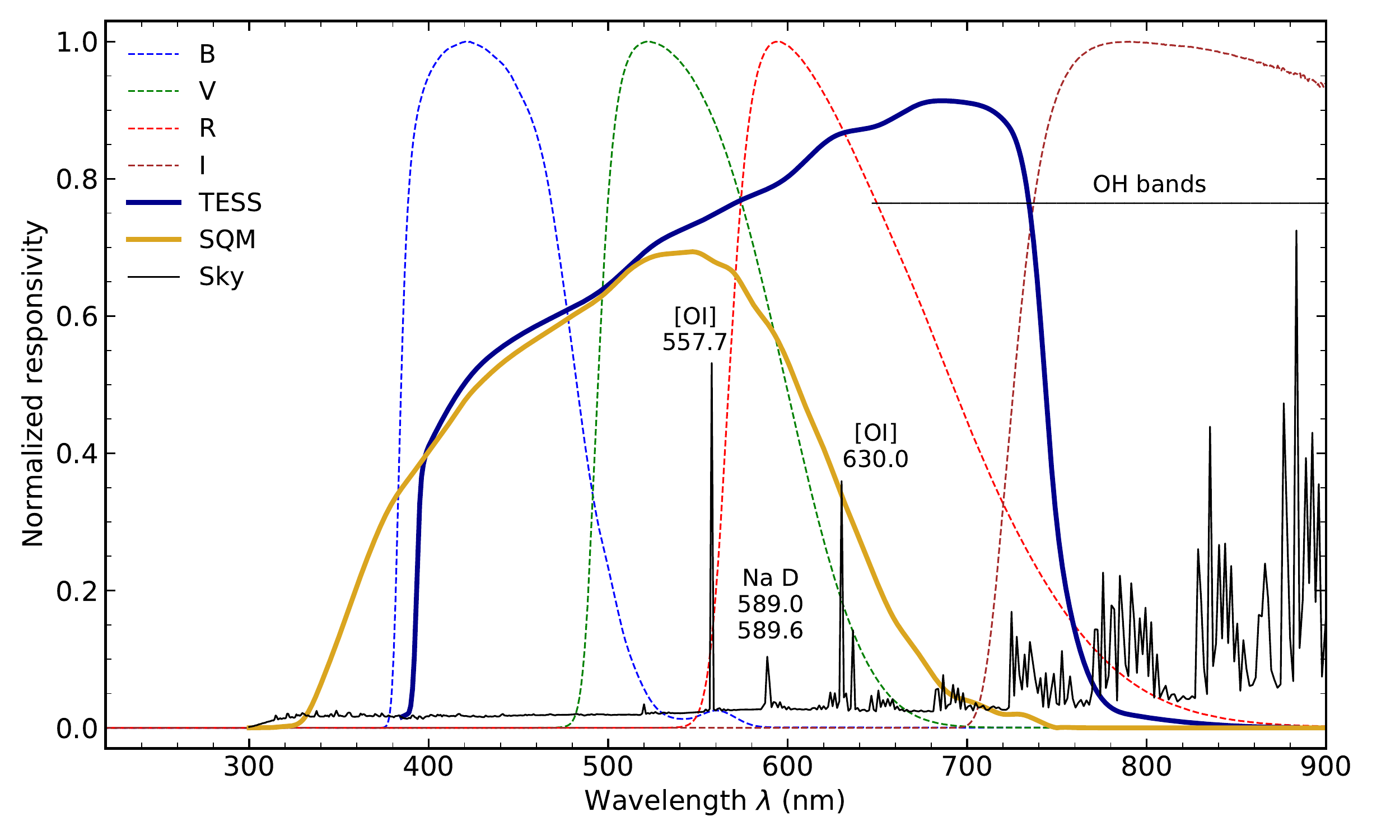}
\caption{Spectral response curve of the TESS photometer. The sensitivity of the TSL237 sensor is limited to the visible range by a dichroic filter, which reaches
redder wavelengths than the SQM. The transmissivity of the Johnson--Cousins \citep{Bessell1990} filters has been included for reference, as well as the night sky spectrum
obtained using the SkyCalc tool \citep{Noll2012}, where the brightest airglow lines are labeled.\label{fig:spectra}}
\end{center}
\end{figure}

The sky brightness detector is a TSL237 photodiode that
converts light to frequency. It is the same sensor used by the
SQM photometer \citep{Cinzano2007}. However, the bandpass of
TESS is more extended to the red range because of the use of a
dichroic filter. Figure \ref{fig:spectra} shows a comparison of the spectral
responses of the SQM and TESS photometers (Johnson--Cousins filters are also included). A typical night sky spectrum
is also shown.

The stability of the measurements was studied in order to
characterize the intrinsic error of the optical system. To do so,
5000 measurements were performed for TESS magnitudes of
21 and 22 mag/arcsec$^2$ (the NSB values expected for the dark
sites included in this work). For each set of 5000 measurements,
a histogram of frequency was obtained (represented in
Figure \ref{fig:instr_TESS}), along with its mean and standard deviation. The
final adopted instrumental error is 0.002 mag/arcsec$^2$ for both
magnitudes. This is an upper limit, because of the nonzero
possibility that it may include the power supply or lamp
variability, external diffuse light, or even electromagnetic
interference. The measurements presented here have been made
in collaboration with Sieltec Canarias S.L. The power source
employed was a PS 2042-20B (Elektro-Automatik), with a
HALOSTAR 50 W 12 V GY6.35 (OSRAM) lamp and an
AvaSpec-ULS3648-UA-25 (AVANTES) spectrometer with a
resolution of 1.1 nm.

Figure \ref{tab:violin} shows the distribution of the standard deviation
grouping the data of each photometer at different time
intervals, from minutes to months. On the minute timescale,
the calculated $\sigma_{\rm TESS}$ is below 0.004 mag/arcsec$^2$, compatible
with the instrumental error obtained above.

\end{document}